\documentclass[11pt,nofootinbib,floatfix,pra]{revtex4}%
\usepackage{amsfonts,amssymb,stmaryrd,latexsym,amsmath,braket}
\usepackage{graphicx,subfigure}
\usepackage[export]{adjustbox}
\usepackage{times}
\usepackage{bbm}
\usepackage{bm}
\usepackage{mathrsfs}
\usepackage{slashed}
\usepackage[version=3]{mhchem}
\usepackage{braket}
\usepackage{hyperref}
\usepackage{verbatim}
\usepackage[utf8]{inputenc}
\usepackage{tikz}
\usepackage[percent]{overpic}

\usetikzlibrary{patterns}

\usepackage{enumitem}

\usepackage{graphicx,subfigure}
\usepackage{times}
\usepackage{slashed}
\usepackage{braket}
\usepackage{verbatim}
\usepackage{multirow}
\usepackage[utf8]{inputenc}
\usepackage{etoolbox}
\usepackage{array}
\newcommand{\PreserveBackslash}[1]{\let\temp=\\#1\let\\=\temp}
\newcolumntype{C}[1]{>{\PreserveBackslash\centering}p{#1}}
\newcolumntype{R}[1]{>{\PreserveBackslash\raggedleft}p{#1}}
\newcolumntype{L}[1]{>{\PreserveBackslash\raggedright}p{#1}}

\usepackage{tabularx}

\makeatletter
\let\orgdescriptionlabel\descriptionlabel
\renewcommand*{\descriptionlabel}[1]{%
  \let\orglabel\label
  \let\label\@gobble
  \phantomsection
  \edef\@currentlabel{#1}%
  \let\label\orglabel
  \orgdescriptionlabel{#1}%
}
\makeatother

\usepackage{color}

\usepackage{makecell}
\definecolor{DarkMidnightBlue}{rgb}{0.0, 0.04, 0.14}

\hypersetup{hidelinks,
breaklinks=true,
colorlinks=true,
linkcolor=blue, %
citecolor=blue, %
urlcolor=DarkMidnightBlue,
bookmarksopen=false,    
pdftitle={Three-nucleon forces on the lattice},
pdfauthor={S. Elhatisari, U.-G. Meissner, Z. Ren}}

\def\affHISKP{\affiliation{Helmholtz-Institut f\"ur Strahlen- und Kernphysik~(Theorie) and Bethe Center for Theoretical Physics, Universit\"at Bonn, D-53115 Bonn, Germany}}

\def\affGIBTU{\affiliation{Faculty of Natural Sciences and Engineering, Gaziantep Islam Science and Technology University, Gaziantep 27010, Turkey}}

\def\affJUELICH{\affiliation{
Institute~for~Advanced~Simulation (IAS-4), Forschungszentrum~J\"{u}lich, D-52425~J\"{u}lich,~Germany}}

\def\affPENG{\affiliation{Peng Huanwu Collaborative Center for Research and Education, International Institute for Interdisciplinary and Frontiers, Beihang University, Beijing 100191, China}}

\def\affSA{\affiliation{King Fahd University of Petroleum and Minerals (KFUPM), 31261 Dhahran, Saudi Arabia}}

\input{figures/interaction_diagrams.tex}
\newcommand{\PWoneSzero}{{}^{1}S_{0}}
\newcommand{\PWthreeSone}{{}^{3}S_{1}}
\newcommand{\PWonePone}{{}^{1}P_{1}}
\newcommand{\PWthreePzero}{{}^{3}P_{0}}
\newcommand{\PWthreePone}{{}^{3}P_{1}}
\newcommand{\PWthreePtwo}{{}^{3}P_{2}}
\newcommand{\PWthreeSoneDone}{{}^{3}S_{1}\text{--}{}^{3}D_{1}}

\csdef{D_1S0}{$\tilde{D}_{\PWoneSzero}$}
\csdef{D_3S1}{$\tilde{D}_{\PWthreeSone}$}

\csdef{D_1S0_01}{$D_{\PWoneSzero}$}          
\csdef{D_3S1_01}{$D_{\PWthreeSone}$}         
\csdef{D_1P1}{$D_{\PWonePone}$}
\csdef{D_3P0}{$D_{\PWthreePzero}$}
\csdef{D_3P1}{$D_{\PWthreePone}$}
\csdef{D_3P2}{$D_{\PWthreePtwo}$}
\csdef{D_3S1_3D1}{$D_{\PWthreeSoneDone}$}    

\csdef{D_PW_generic}{$D^{(d)}_{\alpha}$}
\csdef{V_PW_generic}{$V^{(d)}_{{\rm 3N},\alpha}$}
\csdef{O_PW_generic}{${\cal O}^{\rm PW}_{\alpha}$}

\csdef{V_PW_1S0}{$\tilde{V}^{\rm PW}_{\PWoneSzero}$}
\csdef{V_PW_3S1}{$\tilde{V}^{\rm PW}_{\PWthreeSone}$}
\csdef{V_PW_1S0_01}{$V^{\rm PW}_{\PWoneSzero}$}
\csdef{V_PW_3S1_01}{$V^{\rm PW}_{\PWthreeSone}$}
\csdef{V_PW_1P1}{$V^{\rm PW}_{\PWonePone}$}
\csdef{V_PW_3P0}{$V^{\rm PW}_{\PWthreePzero}$}
\csdef{V_PW_3P1}{$V^{\rm PW}_{\PWthreePone}$}
\csdef{V_PW_3P2}{$V^{\rm PW}_{\PWthreePtwo}$}
\csdef{V_PW_3S1_3D1}{$V^{\rm PW}_{\PWthreeSoneDone}$}

\csdef{D_1S0_nn00}{$D^{\,0,0}_{\PWoneSzero}$}
\csdef{D_3S1_nn00}{$D^{\,0,0}_{\PWthreeSone}$}
\csdef{D_1S0_nn01}{$D^{\,0,1}_{\PWoneSzero}$}
\csdef{D_3S1_nn01}{$D^{\,0,1}_{\PWthreeSone}$}
\csdef{V_PW_1S0_nn01}{$[V^{\rm PW}_{\PWoneSzero}]^{0,1}$}
\csdef{V_PW_3S1_nn01}{$[V^{\rm PW}_{\PWthreeSone}]^{0,1}$}

\csdef{D_1S0_nn02}{$D^{\,0,2}_{\PWoneSzero}$}
\csdef{D_1S0_nn11}{$D^{\,1,1}_{\PWoneSzero}$}
\csdef{D_3S1_nn02}{$D^{\,0,2}_{\PWthreeSone}$}
\csdef{D_3S1_nn11}{$D^{\,1,1}_{\PWthreeSone}$}
\csdef{D_1D2}{$D_{{}^{1}D_{2}}$}
\csdef{D_3D1}{$D_{{}^{3}D_{1}}$}
\csdef{D_3D2}{$D_{{}^{3}D_{2}}$}
\csdef{D_3D3}{$D_{{}^{3}D_{3}}$}
\csdef{D_3P2_3F2}{$D_{{}^{3}P_{2}\text{--}{}^{3}F_{2}}$}

\usepackage{placeins}

\providecommand{\Res}[3]{\csname Result#1#2#3\endcsname}
\providecommand{\Sci}[1]{\csname Science#1\endcsname}
\providecommand{\Fixed}[1]{\csname Fixed#1\endcsname}
\expandafter\def\csname ResultA3rel\endcsname{2.68}
\expandafter\def\csname ResultA3pthree\endcsname{0.327}
\expandafter\def\csname ResultA3pone\endcsname{0.469}
\expandafter\def\csname ResultA3maxdev\endcsname{7.2}
\expandafter\def\csname ResultA3meanerr\endcsname{3.5}
\expandafter\def\csname ResultA3relrange\endcsname{2.68\text{--}2.69}
\expandafter\def\csname ResultA3pthreerange\endcsname{0.324\text{--}0.327}
\expandafter\def\csname ResultA3ponerange\endcsname{0.465\text{--}0.472}
\expandafter\def\csname ScienceA3ResidualMin\endcsname{-0.72}
\expandafter\def\csname ScienceA3ResidualMax\endcsname{+0.19}
\expandafter\def\csname ResultA4rel\endcsname{2.43}
\expandafter\def\csname ResultA4pthree\endcsname{0.167}
\expandafter\def\csname ResultA4pone\endcsname{0.235}
\expandafter\def\csname ResultA4maxdev\endcsname{6.5}
\expandafter\def\csname ResultA4meanerr\endcsname{2.9}
\expandafter\def\csname ResultA4relrange\endcsname{2.41\text{--}2.45}
\expandafter\def\csname ResultA4pthreerange\endcsname{0.163\text{--}0.170}
\expandafter\def\csname ResultA4ponerange\endcsname{0.235\text{--}0.238}
\expandafter\def\csname ScienceA4ResidualMin\endcsname{-0.25}
\expandafter\def\csname ScienceA4ResidualMax\endcsname{+0.39}
\expandafter\def\csname ResultA5rel\endcsname{2.22}
\expandafter\def\csname ResultA5pthree\endcsname{0.174}
\expandafter\def\csname ResultA5pone\endcsname{0.234}
\expandafter\def\csname ResultA5maxdev\endcsname{5.2}
\expandafter\def\csname ResultA5meanerr\endcsname{3.0}
\expandafter\def\csname ResultA5relrange\endcsname{2.21\text{--}2.27}
\expandafter\def\csname ResultA5pthreerange\endcsname{0.173\text{--}0.182}
\expandafter\def\csname ResultA5ponerange\endcsname{0.231\text{--}0.234}
\expandafter\def\csname ScienceA5ResidualMin\endcsname{-0.26}
\expandafter\def\csname ScienceA5ResidualMax\endcsname{+0.38}
\expandafter\def\csname ResultA6rel\endcsname{2.15}
\expandafter\def\csname ResultA6pthree\endcsname{0.160}
\expandafter\def\csname ResultA6pone\endcsname{0.245}
\expandafter\def\csname ResultA6maxdev\endcsname{4.8}
\expandafter\def\csname ResultA6meanerr\endcsname{2.8}
\expandafter\def\csname ResultA6relrange\endcsname{2.10\text{--}2.17}
\expandafter\def\csname ResultA6pthreerange\endcsname{0.160\text{--}0.166}
\expandafter\def\csname ResultA6ponerange\endcsname{0.241\text{--}0.245}
\expandafter\def\csname ScienceA6ResidualMin\endcsname{-0.25}
\expandafter\def\csname ScienceA6ResidualMax\endcsname{+0.41}
\expandafter\def\csname ResultA7rel\endcsname{1.95}
\expandafter\def\csname ResultA7pthree\endcsname{0.164}
\expandafter\def\csname ResultA7pone\endcsname{0.251}
\expandafter\def\csname ResultA7maxdev\endcsname{3.9}
\expandafter\def\csname ResultA7meanerr\endcsname{2.6}
\expandafter\def\csname ResultA7relrange\endcsname{1.80\text{--}1.95}
\expandafter\def\csname ResultA7pthreerange\endcsname{0.156\text{--}0.171}
\expandafter\def\csname ResultA7ponerange\endcsname{0.251\text{--}0.259}
\expandafter\def\csname ScienceA7ResidualMin\endcsname{-0.22}
\expandafter\def\csname ScienceA7ResidualMax\endcsname{+0.44}
\expandafter\def\csname ResultA8rel\endcsname{1.84}
\expandafter\def\csname ResultA8pthree\endcsname{0.165}
\expandafter\def\csname ResultA8pone\endcsname{0.267}
\expandafter\def\csname ResultA8maxdev\endcsname{3.9}
\expandafter\def\csname ResultA8meanerr\endcsname{2.7}
\expandafter\def\csname ResultA8relrange\endcsname{1.83\text{--}1.92}
\expandafter\def\csname ResultA8pthreerange\endcsname{0.161\text{--}0.165}
\expandafter\def\csname ResultA8ponerange\endcsname{0.243\text{--}0.279}
\expandafter\def\csname ScienceA8ResidualMin\endcsname{-0.19}
\expandafter\def\csname ScienceA8ResidualMax\endcsname{+0.48}
\expandafter\def\csname ResultAbaseline\endcsname{21.6}
\expandafter\def\csname ResultB2rel\endcsname{3.83}
\expandafter\def\csname ResultB2pthree\endcsname{0.683}
\expandafter\def\csname ResultB2pone\endcsname{0.445}
\expandafter\def\csname ResultB2maxdev\endcsname{7.3}
\expandafter\def\csname ResultB2meanerr\endcsname{3.8}
\expandafter\def\csname ResultB2relrange\endcsname{3.62\text{--}4.50}
\expandafter\def\csname ResultB2pthreerange\endcsname{0.500\text{--}0.749}
\expandafter\def\csname ResultB2ponerange\endcsname{0.445\text{--}0.461}
\expandafter\def\csname ScienceB2ResidualMin\endcsname{-0.56}
\expandafter\def\csname ScienceB2ResidualMax\endcsname{+0.65}
\expandafter\def\csname ResultB3rel\endcsname{3.78}
\expandafter\def\csname ResultB3pthree\endcsname{0.526}
\expandafter\def\csname ResultB3pone\endcsname{0.442}
\expandafter\def\csname ResultB3maxdev\endcsname{7.3}
\expandafter\def\csname ResultB3meanerr\endcsname{3.3}
\expandafter\def\csname ResultB3relrange\endcsname{3.64\text{--}4.33}
\expandafter\def\csname ResultB3pthreerange\endcsname{0.282\text{--}0.582}
\expandafter\def\csname ResultB3ponerange\endcsname{0.442\text{--}0.491}
\expandafter\def\csname ScienceB3ResidualMin\endcsname{-0.58}
\expandafter\def\csname ScienceB3ResidualMax\endcsname{+0.62}
\expandafter\def\csname ResultB4rel\endcsname{3.08}
\expandafter\def\csname ResultB4pthree\endcsname{0.383}
\expandafter\def\csname ResultB4pone\endcsname{0.552}
\expandafter\def\csname ResultB4maxdev\endcsname{6.3}
\expandafter\def\csname ResultB4meanerr\endcsname{3.1}
\expandafter\def\csname ResultB4relrange\endcsname{3.08\text{--}3.40}
\expandafter\def\csname ResultB4pthreerange\endcsname{0.322\text{--}0.425}
\expandafter\def\csname ResultB4ponerange\endcsname{0.516\text{--}0.552}
\expandafter\def\csname ScienceB4ResidualMin\endcsname{-0.97}
\expandafter\def\csname ScienceB4ResidualMax\endcsname{+0.41}
\expandafter\def\csname ResultB5rel\endcsname{3.03}
\expandafter\def\csname ResultB5pthree\endcsname{0.124}
\expandafter\def\csname ResultB5pone\endcsname{0.395}
\expandafter\def\csname ResultB5maxdev\endcsname{4.9}
\expandafter\def\csname ResultB5meanerr\endcsname{3.5}
\expandafter\def\csname ResultB5relrange\endcsname{2.99\text{--}3.03}
\expandafter\def\csname ResultB5pthreerange\endcsname{0.095\text{--}0.157}
\expandafter\def\csname ResultB5ponerange\endcsname{0.385\text{--}0.406}
\expandafter\def\csname ScienceB5ResidualMin\endcsname{-0.51}
\expandafter\def\csname ScienceB5ResidualMax\endcsname{+0.54}
\expandafter\def\csname ResultB6rel\endcsname{2.89}
\expandafter\def\csname ResultB6pthree\endcsname{0.126}
\expandafter\def\csname ResultB6pone\endcsname{0.406}
\expandafter\def\csname ResultB6maxdev\endcsname{4.7}
\expandafter\def\csname ResultB6meanerr\endcsname{3.3}
\expandafter\def\csname ResultB6relrange\endcsname{2.89\text{--}2.93}
\expandafter\def\csname ResultB6pthreerange\endcsname{0.110\text{--}0.129}
\expandafter\def\csname ResultB6ponerange\endcsname{0.396\text{--}0.409}
\expandafter\def\csname ScienceB6ResidualMin\endcsname{-0.52}
\expandafter\def\csname ScienceB6ResidualMax\endcsname{+0.56}
\expandafter\def\csname ResultB7rel\endcsname{2.92}
\expandafter\def\csname ResultB7pthree\endcsname{0.086}
\expandafter\def\csname ResultB7pone\endcsname{0.402}
\expandafter\def\csname ResultB7maxdev\endcsname{5.3}
\expandafter\def\csname ResultB7meanerr\endcsname{2.9}
\expandafter\def\csname ResultB7relrange\endcsname{2.88\text{--}2.92}
\expandafter\def\csname ResultB7pthreerange\endcsname{0.077\text{--}0.104}
\expandafter\def\csname ResultB7ponerange\endcsname{0.390\text{--}0.407}
\expandafter\def\csname ScienceB7ResidualMin\endcsname{-0.48}
\expandafter\def\csname ScienceB7ResidualMax\endcsname{+0.58}
\expandafter\def\csname ResultB8rel\endcsname{2.59}
\expandafter\def\csname ResultB8pthree\endcsname{0.113}
\expandafter\def\csname ResultB8pone\endcsname{0.412}
\expandafter\def\csname ResultB8maxdev\endcsname{5.2}
\expandafter\def\csname ResultB8meanerr\endcsname{2.9}
\expandafter\def\csname ResultB8relrange\endcsname{2.50\text{--}2.71}
\expandafter\def\csname ResultB8pthreerange\endcsname{0.112\text{--}0.129}
\expandafter\def\csname ResultB8ponerange\endcsname{0.400\text{--}0.413}
\expandafter\def\csname ScienceB8ResidualMin\endcsname{-0.50}
\expandafter\def\csname ScienceB8ResidualMax\endcsname{+0.58}
\expandafter\def\csname ResultBbaseline\endcsname{21.6}
\expandafter\def\csname SciencePthreeReduction\endcsname{-4}
\expandafter\def\csname MainFiveRel\endcsname{2.22}
\expandafter\def\csname MainFivePthree\endcsname{0.174}
\expandafter\def\csname FixedSpacing\endcsname{1.32}
\expandafter\def\csname FixedMass\endcsname{938.92}
\expandafter\def\csname FixedAxial\endcsname{1.287}
\expandafter\def\csname FixedFpi\endcsname{92.2}
\expandafter\def\csname FixedMpi\endcsname{134.98}
\expandafter\def\csname FixedPionCutoff\endcsname{300}
\expandafter\def\csname FixedChiralScale\endcsname{700}
\expandafter\def\csname FixedLocalSmearing\endcsname{0.07}
\expandafter\def\csname FixedCutoff\endcsname{470}
\expandafter\def\csname FixedPionLength\endcsname{1.46}
\expandafter\def\csname FixedLocalRange0\endcsname{0.00}
\expandafter\def\csname FixedLocalRange1\endcsname{1.32}
\expandafter\def\csname FixedLocalRange2\endcsname{1.87}
\expandafter\def\csname FixedLocalRange3\endcsname{2.29}
\expandafter\def\csname FixedRadius011\endcsname{0.62}
\expandafter\def\csname FixedRadius022\endcsname{0.88}
\expandafter\def\csname FixedRadius033\endcsname{1.08}
\expandafter\def\csname FixedRadius044\endcsname{1.24}
\expandafter\def\csname FixedRadius112\endcsname{0.88}
\expandafter\def\csname FixedRadius114\endcsname{1.08}
\expandafter\def\csname FixedRadius123\endcsname{1.08}
\expandafter\def\csname FixedRadius125\endcsname{1.24}
\expandafter\def\csname FixedRadius145\endcsname{1.39}
\expandafter\def\csname FixedRadius222\endcsname{1.08}
\expandafter\def\csname FixedRadius224\endcsname{1.24}
\expandafter\def\csname FixedRadius235\endcsname{1.39}
\expandafter\def\csname FixedRadius334\endcsname{1.39}

\begin{document}

\title{Constraining three-nucleon forces in nuclear lattice effective field theory}

\author{Serdar Elhatisari}
\email{selhatisari@gmail.com}
\affGIBTU
\affSA

\author{Ulf-G. Mei{\ss}ner}
\email{meissner@hiskp.uni-bonn.de}
\affHISKP
\affJUELICH
\affPENG

\author{Zhengxue Ren}
\email{zxren@nankai.edu.cn}
\affiliation{School of Physics, Nankai University, Tianjin 300071, China}

\author{Shuang Zhang}
\email{shu.zhang@fz-juelich.de}
\affJUELICH

\begin{abstract}
We study short-range three-nucleon forces (3NFs) in nuclear lattice effective field theory with a chiral two-nucleon interaction at N${}^3$LO. The one-pion-exchange ($c_D$) and contact ($c_E$) terms of the leading 3NF are decomposed into locally and non-locally smeared components and supplemented by partial-wave-resolved pair contacts coupled to the density of a third nucleon. We determine their coefficients from nuclear binding energies for nuclei in the mass range from $A=3$ to $A=40$ and neutron--alpha finite-volume energies. This allows to work out the minimal three-nucleon
force that can be used for future precision studies in NLEFT.
\end{abstract}

\maketitle

\section{Introduction and extended overview of three-nucleon forces in {\em ab initio} calculations}

The ultimate goal of \textit{ab initio} nuclear theory is to describe the properties of atomic nuclei and the saturation properties of nuclear matter starting from the effective degrees of freedom and microscopic interactions derived from the underlying theory. Toward this goal, a variety of \textit{ab initio} approaches have made remarkable progress over the past two decades, yielding accurate results for nuclear structure across broad regions of the nuclear chart~\cite{Drischler:2017wtt,Lonardoni:2017hgs,Morris:2017vxi,Piarulli:2017dwd,Soma:2019bso,Gysbers:2019uyb,Maris:2020qne,Jiang:2020the,Heinz:2021xir,Abe:2021sky,Wirth:2021pij,Sarma:2022pml,LENPIC:2022cyu,Kondo:2023lty,Arthuis:2024mnl}. A major reason why these approaches have been able to address such broad nuclear-structure and nuclear-matter problems is the availability of systematic and improvable nuclear forces derived within chiral effective field theory (EFT), constrained by the symmetries of QCD~\cite{Weinberg:1990,Weinberg:1991,Epelbaum:2002ji,Entem:2003ft,Epelbaum:2008ga,Machleidt:2011,Gezerlis:2013ipa,Epelbaum:2014efa,Epelbaum:2014sza,Piarulli:2014bda,PhysRevC.91.051301,Piarulli:2016vel,Piarulli:2017dwd,Entem:2017,Reinert:2017usi,Ekstrom:2017koy,Epelbaum:2019kcf,Jiang:2020the,HUTHER2020135651}. These high-precision nuclear forces have been constructed at different orders up to N${}^4$LO, and beyond, in the chiral expansion. Their essential differences include the choice of regularization scheme, the strategy used to determine the low-energy constants (LECs) of the short-range two-nucleon forces (2NFs) and three-nucleon forces (3NFs), the treatment of explicit excited nucleonic degrees of freedom such as the $\Delta$ isobar, and the methods used to estimate theoretical uncertainties associated with the truncation of the chiral expansion.

The general procedure of controlling chiral interactions is to regulate them at a given order using a local or non-local regulator in momentum or configuration space for a finite momentum as a cutoff. For the choice of the cutoff, all the high-precision nuclear forces use the strategy that the cutoff should be only varied in a range (around the expected breakdown scale) where the accuracy of the reproduction of the experimental data shows convergence~\cite{Lepage:1997cs}. Therefore, the essential difference between the regularization schemes is the degrees of locality and non-locality of the chosen regulator. The chiral interactions of Ref.~\cite{Epelbaum:2014sza} are formulated up to N${}^4$LO where the long-range interactions are regulated using a local coordinate-space regulator, while the short-range interactions are regulated using a non-local Gaussian regulator in momentum space. This semi-local regularization approach was improved in Ref.~\cite{Reinert:2017usi}, and the long-range interactions were regulated in momentum space, which makes  the interactions more suitable for implementation in many-body systems. On the other hand, both the short-range and the long-range parts of chiral interactions of Ref.~\cite{Entem:2017} are regulated using non-local exponential regulators in momentum space and chiral interactions up to N${}^4$LO are available, and an additional spectral function regularization~\cite{Epelbaum:2003gr,Epelbaum:2003xx} is used for the short-range part of the long-range two-pion exchange (TPE) potential. In addition,  by removing all sources of non-localities, local chiral interactions at N${}^2$LO regulated in configuration space were introduced in Ref.~\cite{Gezerlis:2013ipa}. Interactions including the $\Delta$-isobar in intermediate states of the long-range TPE potential through N${}^2$LO and contact interactions through N${}^3$LO were formulated in a minimally non-local coordinate-space form~\cite{Piarulli:2014bda} and subsequently in a local form~\cite{Piarulli:2016vel}. The chiral interactions including explicit $\Delta$-isobar degrees of freedom are also formulated up to N${}^2$LO using non-local regulator for all the interactions in momentum space and using a spectral function regularization~\cite{Epelbaum:2003gr,Epelbaum:2004fk} for the short-range part of the long-range two-pion exchange (TPE) potential~\cite{Jiang:2020the,Ekstrom:2017koy}. More recently, a new formulation based on the gradient flow was proposed~\cite{Krebs:2023ljo,Krebs:2023gge}, which provides a framework for deriving consistently regularized two- and three-nucleon forces.

\begin{figure}[tb]
\centering\includegraphics[width=0.5\textwidth]{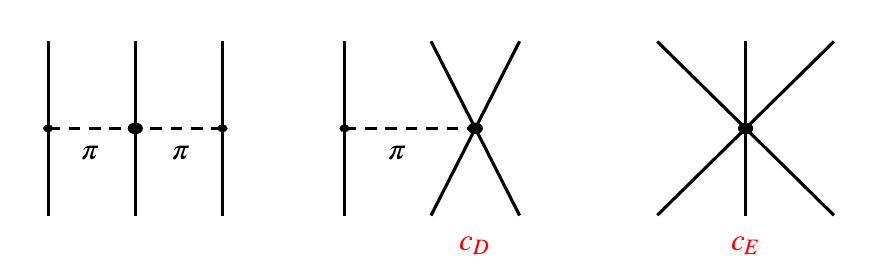}
\caption{Leading 3NFs at N${}^2$LO in the chiral expansion in a theory without delta isobars. These are the two-pion exchange term (left),
the one-pion exchange term with the LEC $c_D$ (middle) and purely short-range three-nucleon contact term with the LEC $c_E$ shown on the right.  At N${}^3$LO there are additional parameter-free pion-exchange, short-range and relativistic corrections to the 3NF.}
\label{fig:threeN}
\end{figure}

The common feature of these high-precision formulations of the chiral interactions is to provide various rigorous and reliable tools to address the quantification of the theoretical uncertainties from the truncation of the chiral expansion and regulator dependence, and the statistical uncertainties in the determination of the LECs, which is crucial to assess the predictive power of \textit{ab initio} calculations of atomic nuclei and nuclear matter. These families of chiral interactions include interactions beyond 2NFs. Although the incorporation of 2NFs at N${}^3$LO into \textit{ab initio} calculations was firmly established two decades ago~\cite{Navrtil:2004qx,Gour:2005dm} and high-precision 2NFs at N${}^4$LO have been already available for approximately a decade, the incorporation of 3NFs remains a significant challenge within the realm of many-body nuclear theory, see Ref.~\cite{Hebeler:2020ocj} for a comprehensive review. In chiral EFT without explicit $\Delta$ degrees of freedom, the leading nonvanishing 3NFs appear at N${}^2$LO. As shown in Fig.~\ref{fig:threeN}, these include the long-range two-pion exchange (TPE) interaction which is parameterized by the subleading pion-nucleon couplings $c_1$, $c_3$ and $c_4$, which can be constrained by pion-nucleon scattering, including Roy-Steiner analyses~\cite{Hoferichter:2015tha,Hoferichter:2015hva}, the long-range one-pion exchange (OPE) linked to the short-range interaction of two nucleons with the corresponding unknown coupling $c_{D}$, and the purely three-body short-range interaction with the unknown coupling $c_{E}$. At N${}^3$LO the 3NF receives corrections from the two-pion-exchange, two-pion--one-pion-exchange and ring topologies as well as from short-range and relativistic $1/m_N$ terms~\cite{Ishikawa:2007zz,Bernard:2007sp,Bernard:2011zr} with no new independent 3NF LECs, so that up to N${}^3$LO the 3NF depends, apart from the previously determined pion--nucleon and NN LECs, only on $c_D$ and $c_E$. Genuinely new three-nucleon contact LECs first appear at N${}^4$LO~\cite{Girlanda:2011fh}, where also the long- and intermediate-range terms, fixed by pion--nucleon input, are known~\cite{Krebs:2012yv,Krebs:2013kha}. Consequently, up to N${}^3$LO the determination of the LECs of the 3NFs essentially reduces to a two-parameter problem, and the properties of 3N systems are expected to serve as a particularly natural means of constraining these LECs. Early derivations and analyses of chiral 3NFs were given in Refs.~\cite{vanKolck:1994yi,Friar:1998zt}. In Ref.~\cite{Epelbaum:2002vt}, the 3NFs at N${}^2$LO were regulated non-locally in momentum space and the LECs were determined by using neutron-deuteron doublet scattering length and ${}^3$H binding energy, and the ground state energy of ${}^4$He was predicted. Shortly after, the \textit{ab initio} no-core shell model (NCSM) approach used the 2NFs at N${}^3$LO of Ref.~\cite{Entem:2003ft} and the 3NFs at N${}^2$LO of Ref.~\cite{Epelbaum:2002vt} by constraining the 3NFs LECs to the ${}^3$H and  ${}^4$He binding energies, and the excited states of ${}^6$Li~\cite{Barrett:2004qsq} and the spectrum of ${}^7$Li~\cite{Nogga:2005hp} were predicted. These NCSM calculations were extended to the structure properties of nuclei ${}^4$He, ${}^6$Li, ${}^{10,11}$B and ${}^{12,13}$C~\cite{Navratil:2007we}. In this work, the 3NF LECs at N${}^2$LO were constrained by the binding energies of the $A=3$ nuclei, the ${}^4$He radius and selected properties of ${}^6$Li, ${}^{10}$B and ${}^{12}$C. In 2009, Ref.~\cite{Gazit:2008ma} constrained the 3NF LECs using the binding energies of the $A=3$ nuclei and, for the first time, the tritium $\beta$-decay matrix element. Using the locally regulated 3NF, the radii of the $A=3$ nuclei and properties of ${}^4$He were predicted. The local N${}^2$LO 3NF formulated in 2007 in Ref.~\cite{Navratil:2007zn} was used in the \textit{ab initio} NCSM calculations of $A=7,8$ nuclei in Ref.~\cite{Maris:2012bt}, with the LECs constrained by the $A=3$ binding energies and tritium $\beta$ decay. In 2011, this local 3NF and the same calibration approach were used together with chiral 2NFs at N${}^3$LO, with both interactions evolved through the similarity renormalization group (SRG), in \textit{ab initio} NCSM calculations of ${}^4$He, ${}^6$Li, ${}^{12}$C and ${}^{16}$O~\cite{Roth:2011ar}. The same calibration approach was also adopted for the baseline local interaction in Coupled-Cluster (CC) calculations of the Equation of State (EoS) of nucleonic matter at N${}^2$LO~\cite{Hagen:2013yba}. In Ref.~\cite{Roth:2011vt}, the 2NFs and 3NFs were evolved consistently through the SRG. In addition to the original interaction, a local 3NF with a cutoff of 400~MeV was used, keeping $c_D$ fixed and readjusting $c_E$ to the ${}^4$He binding energy. These chiral interactions were used to calculate the ground state energies of $^4$He, $^{16}$O, $^{40}$Ca using the importance-truncated NCSM (IT-NCSM) and $^{16,24}$O, $^{40,48}$Ca using the CC method~\cite{Roth:2011vt}, and $^{13,27}$N, $^{14,28}$O and $^{15,29}$F using the self-consistent Green’s function theory (SCGF) with the lower-cutoff interaction~\cite{Cipollone:2013zma}. The latter interaction was also used in \textit{ab initio} Bogoliubov many-body perturbation theory (BMBPT) calculations of the ground-state energies of nuclei $^{16-26}$O, $^{36-60}$Ca and $^{50-78}$Ni~\cite{Tichai:2018}. In Ref.~\cite{Soma:2019bso}, based on the approaches of Ref.~\cite{Cipollone:2013zma} and defining both locally and non-locally regulated 3NFs, a variant of the chiral interaction was introduced, and using NCSM and SCGF methods the binding energies of light and medium-mass nuclei were reproduced. Another approach used to determine the  LECs $c_{E}$ and $c_{D}$  was to constrain the binding energy of ${}^4$He and the low-energy scattering P--wave phase shifts of neutron-${}^4$He to improve the predictive power of the calculations for neutron-rich nuclei and neutron matter. This was done for the local 3NFs at N${}^2$LO, and using locally regulated 2NFs at N${}^2$LO~\cite{Gezerlis:2013ipa} the auxiliary field diffusion quantum Monte Carlo (AFDMC) calculations were performed for light nuclei binding energies, neutron-${}^4$He and the EoS of nucleonic matter~\cite{Lynn:2015jua}. In Ref.~\cite{Epelbaum:2009zsa}, lattice simulations using auxiliary field quantum Monte Carlo projection were extended to N${}^2$LO, including the leading chiral 3NFs~\cite{Epelbaum:2002ji,Epelbaum:2002vt}. In this work, in the determination of the 3NFs LECs the same approach followed by Ref.~\cite{Epelbaum:2002vt} was used and the ground state energy of ${}^4$He was predicted. The lattice simulations later were extended to the calculations of the ground state energies of ${}^3$H, ${}^4$He, ${}^{6}$Li and ${}^{12}$C~\cite{Epelbaum:2009pd}, the first \textit{ab initio} calculation of the Hoyle State~\cite{Hoyle:1954zz,Epelbaum:2011md,Epelbaum:2012qn}, the spectrum and structure calculations of $^{16}$O~\cite{Epelbaum:2013paa}, the ground state energies of alpha nuclei from ${}^4$He to ${}^{28}$Si~\cite{Lahde:2013uqa}, and the first \textit{ab initio} calculation of ${}^4$He-${}^4$He scattering~\cite{Elhatisari:2015iga}. In the framework of lattice simulations using locally regulated chiral interactions, however, reproducing the ground state energies of nuclei beyond $^{16}$O remained problematic due to an overbinding problem.

In Refs.~\cite{Bogner:2009un,Hebeler:2010xb} an alternative approach was followed to determine the LECs of the 3NFs at N${}^2$LO by fitting to the binding energy of  ${}^3$H and the charge radius of ${}^4$He to improve the predictive power of the calculations of nuclear matter. In this work, the short-range 3NFs were regulated non-locally in momentum space. Using a combined chiral 2NFs at N${}^3$LO and 3NFs at N${}^2$LO, symmetric nuclear matter~\cite{Bogner:2009un,Hebeler:2010xb}, asymmetric nuclear matter~\cite{Drischler:2013iza,Drischler:2015eba}, ground-state energies of exotic calcium isotopes  ${}^{48-58}$Ca~\cite{Wienholtz:2013nya}, ground-state and excited-state properties of all \textit{sd}-shell nuclei with the neutron and proton numbers ranging from 8 to 20~\cite{Simonis:2015vja}, and ground-state energies and charge radii of open- and closed-shell nuclei up to $A=78$~\cite{Simonis:2017dny} were studied. These interactions gave a good description of many ground-state and separation energies, while a simultaneous description of binding energies and charge radii remained challenging~\cite{Simonis:2017dny}. In Ref.~\cite{Logoteta:2019zuc} the chiral interactions of Ref.~\cite{Entem:2017} were used and the LECs of the 3NFs were adjusted by fitting to properties of light nuclei with $A = 3$. Then by employing different regularization schemes for the 3NFs, a good description for some properties of nuclear matter was obtained. In a separate development, the LENPIC collaboration determined the 3NF LECs from the triton binding energy and the differential cross-section minimum in elastic nucleon-deuteron scattering~\cite{LENPIC:2018ewt}, and in this work the ground state and excited state energies of light nuclei with $A=3$--$16$ were predicted. Using the semi-local regularization approach of Ref.~\cite{Reinert:2017usi} for 2NFs and 3NFs at N${}^2$LO and determining the LECs of the 3NFs from the triton binding energy and the differential cross section minimum in elastic nucleon-deuteron scattering, Ref.~\cite{Maris:2020qne} performed predictive calculations for selected observables in elastic nucleon-deuteron scattering and in the deuteron breakup reactions, for properties of the $A=3$ and $A=4$ nuclei, and for spectra of
p-shell nuclei up to $A=16$. Subsequently, 2NFs through N${}^4$LO$^{+}$, combined with the N${}^2$LO 3NF, were used to calculate binding energies and radii in the oxygen and calcium isotope chains~\cite{LENPIC:2022cyu}. 

\begin{table}[h!]
\caption{\label{tab:lightnuclei} Constraints on 3NFs in various approaches and the resulting predictions. BE: binding energy, $R_{\rm c}$: charge radius, EoS: equation of state.}
\scriptsize
\centering
\begin{tabular}{|>{\centering\arraybackslash}m{3.7cm}||>{\centering\arraybackslash}m{5.1cm}|m{6.4cm}|}
\hline\hline
Work & Constraints & Predictions \\
\hline\hline
NCSM, Barrett \textit{et al.}~\cite{Barrett:2004qsq} &
BE of ${}^3$H and ${}^4$He &
Excited states of ${}^6$Li\\ \hline
NCSM, Nogga \textit{et al.}~\cite{Nogga:2005hp} &
BE of ${}^3$H and ${}^4$He &
Spectrum of ${}^7$Li\\ \hline
NCSM, Maris \textit{et al.}~\cite{Maris:2012bt} &
BE of ${}^3$H and ${}^3$H $\beta$ decay &
Structures of $A= 7, 8$ \\ \hline
NCSM, Roth \textit{et al.}~\cite{Roth:2011ar} &
BE of ${}^3$H and ${}^3$H $\beta$ decay &
${}^4$He, ${}^6$Li, ${}^{12}$C and ${}^{16}$O \\ \hline
CC, Hagen \textit{et al.}~\cite{Hagen:2013yba} &
BE of ${}^3$H and ${}^3$H $\beta$ decay &
EoS of nucleonic matter \\ \hline
BMBPT, Tichai \textit{et al.}~\cite{Tichai:2018} &
BE of ${}^4$He, with $c_D$ fixed &
BE of $^{16-26}$O, $^{36-60}$Ca and $^{50-78}$Ni \\ \hline
MBPT/IM-SRG, Bogner~\textit{et al.}~\cite{Bogner:2009un}, Hebeler~\textit{et al.}~\cite{Hebeler:2010xb}, Drischler~\textit{et al.}~\cite{Drischler:2013iza,Drischler:2015eba}, Wienholtz~\textit{et al.}~\cite{Wienholtz:2013nya}, Simonis~\textit{et al.}~\cite{Simonis:2015vja,Simonis:2017dny}
&
BE of $^{3}$H and $R_{\rm c}$ of $^{4}$He   
&
symmetric nuclear matter, asymmetric nuclear matter, BE of ${}^{48-58}$Ca, spectrum of \textit{sd}-shell nuclei with the neutron and proton numbers ranging from 8 to 20, BE and $R_{\rm c}$ of open- and closed-shell nuclei up to $A=78$
\\\hline
IT-NCSM, Roth \textit{et al.}~\cite{Roth:2011vt} &
$A=3$ BE and ${}^3$H $\beta$ decay; ${}^4$He BE at fixed $c_D$ for the lower-cutoff variant &
BE of $^{16}$O and $^{40}$Ca; ${}^4$He BE for the original interaction \\ \hline
CC, Roth \textit{et al.}~\cite{Roth:2011vt} &
$A=3$ BE and ${}^3$H $\beta$ decay; ${}^4$He BE at fixed $c_D$ for the lower-cutoff variant &
BE of $^{16,24}$O, $^{40,48}$Ca \\ \hline
SCGF, Cipollone \textit{et al.}~\cite{Cipollone:2013zma} &
BE of ${}^4$He, with $c_D$ fixed &
BE of $^{13,27}$N, $^{14,28}$O and $^{15,29}$F\\ \hline
AFDMC, Lynn~\textit{et al.}~\cite{Lynn:2015jua} &
BE of ${}^4$He and n-$^4$He P-wave phase shifts
& EoS of nucleonic matter \\\hline
NCCI/IM-NCSM/IM-SRG, Epelbaum~\textit{et al.}~\cite{LENPIC:2018ewt}, Maris~\textit{et al.}~\cite{Maris:2020qne,LENPIC:2022cyu} &
BE of ${}^{3}$H and the minimum of the elastic $Nd$ differential cross section at $E_N = 70$~MeV (the $nd$ doublet scattering length being strongly correlated with the ${}^{3}$H BE)  
&
elastic $Nd$ scattering observables and deuteron breakup, properties of the $A=3$ and $A=4$ nuclei, spectra of p-shell nuclei up to $A=16$, BE and $R_{\rm c}$ of the oxygen and calcium isotope chains
\\\hline
NCSM, Navratil \textit{et al.}~\cite{Navratil:2007we} &
$A=3$ BE, ${}^4$He radius and selected properties of ${}^6$Li, ${}^{10}$B, ${}^{12}$C &
Other properties of ${}^4$He, ${}^6$Li, ${}^{10,11}$B and ${}^{12,13}$C\\ \hline
NCSM, Carlsson~\textit{et al.}~\cite{PhysRevX.6.011019} &
simultaneous fit of all $NN$ and 3N LECs to $\pi N$ and $NN$ scattering data, BE and radii of ${}^{2}$H, ${}^{3}$H, ${}^{3}$He and the ${}^{3}$H $\beta$-decay half-life (N${}^2$LO$_{\rm sim}$ family)
& BE and radius of ${}^{4}$He with propagated statistical and $\chi$EFT-truncation uncertainties \\ \hline
CC, Ekström~\textit{et al.}~\cite{PhysRevC.91.051301}, Hagen~\textit{et al.}~\cite{Hagen:2015yea,Hagen:2016uwj} &
$NN$ scattering data, BE of ${}^{3}$H, ${}^{3,4}$He, ${}^{14}$C, ${}^{16,22,24,25}$O and $R_{\rm c}$ of ${}^{3}$H, ${}^{3,4}$He, ${}^{14}$C, ${}^{16}$O (N${}^2$LO$_{\rm sat}$)  
& $R_{\rm c}$ and BE of nuclei up to $^{40}$Ca, symmetric nuclear matter, neutron skin of $^{48}$Ca, structure of $^{78}$Ni \\ \hline
CC, Ekström~\textit{et al.}~\cite{Ekstrom:2017koy} &
BE and $R_{\rm c}$ of ${}^{4}$He ($\Delta$-full and $\Delta$-less N${}^2$LO)
& BE, $R_{\rm c}$ and neutron skins of ${}^{8}$He, ${}^{16,22,24}$O and ${}^{40,48}$Ca,  saturation of symmetric nuclear matter \\ \hline
NCSM, IM-SRG, IM-NCSM, H\"uther~\textit{et al.}~\cite{HUTHER2020135651} &  
BE of $^{3}$H and $^{16}$O
&
$R_{\rm c}$ and BE of ${}^{4}$He, ${}^{14-26}$O, ${}^{36-52}$Ca and ${}^{48-78}$Ni (excluding the fitted ${}^{16}$O BE), the spectrum of ${}^{7}$Li, ${}^{8}$Be, ${}^{9}$Be and ${}^{10}$B
\\ \hline
CC, Jiang~\textit{et al.}~\cite{Jiang:2020the} &
$NN$ scattering data, $A\le 4$ properties, nuclear-matter saturation, symmetry energy and its slope
&
BE and $R_{\rm c}$ of nuclei from $A = 16$ to $A=132$, nuclear-matter EoS away from the fitted constraints, selected excited states \\ \hline
\hline
\end{tabular}
\end{table}

\begin{table}[h!]
\caption{\label{tab:lightnuclei2} Continuation of Table~\ref{tab:lightnuclei}. Notation as in Table~\ref{tab:lightnuclei}.}
\scriptsize
\centering
\begin{tabular}{|>{\centering\arraybackslash}m{3.7cm}||>{\centering\arraybackslash}m{5.1cm}|m{6.4cm}|}
\hline\hline
Work & Constraints & Predictions \\
\hline\hline
GFMC/HH, Piarulli~\textit{et al.}~\cite{Piarulli:2017dwd}, Baroni~\textit{et al.}~\cite{Baroni:2018fdn} &
BE of ${}^{3}$H and $nd$ doublet scattering length (NV2+3); BE of ${}^{3}$H and ${}^{3}$He and ${}^{3}$H $\beta$ decay (NV2+3$^{*}$)
&
ground and excited states of nuclei with $A\le 12$ \\ \hline
AFDMC, Lonardoni~\textit{et al.}~\cite{Lonardoni:2017hgs,Lonardoni:2018nob,Lonardoni:2019ypg} &
BE of ${}^{4}$He and n-${}^{4}$He P-wave phase shifts
&
BE, $R_{\rm c}$ and charge form factors for $3\le A\le16$, EoS of nuclear and neutron-star matter \\ \hline
MBPT/IM-SRG, Drischler~\textit{et al.}~\cite{Drischler:2017wtt}, Hoppe~\textit{et al.}~\cite{Hoppe:2019uyw} &
BE of ${}^{3}$H and the saturation point of symmetric nuclear matter
&
nuclear-matter EoS, BE and $R_{\rm c}$ of closed-shell nuclei \\ \hline
IM-SRG, Stroberg~\textit{et al.}~\cite{Stroberg:2019bch}, Miyagi~\textit{et al.}~\cite{Miyagi:2021pdc} &
BE of ${}^{3}$H and $R_{\rm c}$ of ${}^{4}$He [1.8/2.0~(EM)~\cite{Hebeler:2010xb}]
&
BE, separation energies and drip lines from He to Fe; BE of ${}^{132}$Sn and ${}^{208}$Pb \\ \hline
IM-SRG, Arthuis~\textit{et al.}~\cite{Arthuis:2024mnl} &
BE of ${}^{3}$H, BE and $R_{\rm c}$ of ${}^{16}$O
&
BE and $R_{\rm c}$ of other nuclei up to ${}^{208}$Pb, density distributions and neutron skins \\ \hline
IM-SRG, Hu~\textit{et al.}~\cite{Hu:2026qlt} &
BE of ${}^{3}$H and ${}^{16}$O
&
BE and $R_{\rm c}$ of nuclei up to ${}^{132}$Sn \\ \hline
CC/IM-SRG, Hu~\textit{et al.}~\cite{Hu:2025cjl} &
NN phase shifts and $^1$S$_0$ effective-range parameters; Bound-state observables of $^2$H and $^4$He;
BE and $R_c$ of $^{16}$O;
&
BE and $R_{\rm c}$ of nuclei up to ${}^{208}$Pb; EOS of symmetric nuclear matter and pure neutron matter. \\ \hline
NCSM, Wesolowski~\textit{et al.}~\cite{Wesolowski:2021cni}, Dj\"arv~\textit{et al.}~\cite{Djarv:2021hcj} &
BE of ${}^{3}$H and ${}^{4}$He, radius of ${}^{4}$He and ${}^{3}$H $\beta$ decay
&
BE, separation energies and spectra of $A=6$ nuclei \\ \hline
Faddeev/AFDMC, Curry~\textit{et al.}~\cite{Curry:2025pna} &
BE and $\beta$ decay of ${}^{3}$H, with or without $R_{\rm c}$ of ${}^{4}$He
&
BE of ${}^{3,4,6}$He \\ \hline
CC/IM-SRG, Hu~\textit{et al.}~\cite{Hu:2021trw} &
$NN$ scattering and deuteron properties; BE and radii of ${}^{3}$H, ${}^{4}$He, ${}^{16}$O and ${}^{48}$Ca; $2^+$ energy of ${}^{48}$Ca
&
BE, radius, neutron skin and dipole polarizability of ${}^{208}$Pb, symmetry energy and its slope \\ \hline
CC, Kondo~\textit{et al.}~\cite{Kondo:2023lty} &
$NN$ scattering, properties of ${}^{2,3}$H, ${}^{4}$He and ${}^{16}$O, energies of ${}^{22,24,25}$O
&
energies of ${}^{27,28}$O \\ \hline
CC, Sun~\textit{et al.}~\cite{Sun:2024iht} &
Constraints of Ref.~\cite{Kondo:2023lty}, $2^+,4^+$ excitation energies of ${}^{24}$Ne
&
rotational energies and electromagnetic transitions in other Ne isotopes \\ \hline
\hline
\end{tabular}
\end{table}

In the meantime, the chiral interactions up to N${}^2$LO with and without explicit $\Delta$-isobar were constructed by using the standard non-local regulators in momentum space for both the 2NFs and 3NFs, and the LECs of the 3NFs were determined by fitting to the binding energy and the charge radius of ${}^4$He~\cite{Ekstrom:2017koy}. For earlier work on including the decuplet, see e.g. \cite{vanKolck:1994yi,Krebs:2007rh,Epelbaum:2007sq}.  Earlier, Ref.~\cite{Otsuka:2009cs} demonstrated the role of three-nucleon forces, including contributions from intermediate $\Delta$ excitations, in explaining the oxygen drip line. The interactions of Ref.~\cite{Ekstrom:2017koy} were employed to compute the binding energies, charge radii, and neutron skin for ${}^8$He, ${}^{16,22,24}$O, and ${}^{40,48}$Ca. For the interactions considered there, the $\Delta$-less calculations gave nuclei unstable against breakup into $\alpha$ particles, whereas including the $\Delta$ improved the binding energies and charge radii. In Ref.~\cite{Piarulli:2017dwd}, using chiral interactions introduced in Ref.~\cite{Piarulli:2016vel} and fitting 3NFs with $\Delta$-intermediate states to the ground-state energy of ${}^3$H and the central value of the neutron-deuteron doublet scattering length, the ground states and excited states of ${}^{4,6,8}$He, ${}^{6,7,8,9}$Li, ${}^{8,9,10}$Be, ${}^{10,11}$B and ${}^{12}$C were predicted with accuracy less than 2\%. Ref.~\cite{Logoteta:2016nzc} used chiral interactions introduced in Ref.~\cite{Piarulli:2016vel} and constructed several interaction models depending on how the $c_{E}$ and $c_{D}$ LECs are determined, and the properties of nuclear matter were investigated. These models included fitting to the binding energy of light nuclei with $A = 3$,  the binding energy of light nuclei with $A = 3$ as well as ${}^3$H $\beta$ decay, and the saturation properties of symmetric nuclear matter. 

In Ref.~\cite{PhysRevC.91.051301}, the 2N and 3N LECs were optimized simultaneously to low-energy nucleon-nucleon (NN) scattering data, binding energies of ${}^{3}$H, ${}^{3,4}$He, ${}^{14}$C and ${}^{16,22,24,25}$O, and charge radii of ${}^{3}$H, ${}^{3,4}$He, ${}^{14}$C and ${}^{16}$O, yielding the N${}^2$LO$_{\rm sat}$ interaction. A subsequent study examined simultaneous optimization order by order, including statistical and theoretical uncertainties~\cite{PhysRevX.6.011019}. In N${}^2$LO$_{\rm sat}$, the short-range parts are regulated using a non-local regulator in momentum space, while the long-range parts  (two-pion exchange) are regularized by using spectral function regularization~\cite{Epelbaum:2003gr,Epelbaum:2005pn}. \textit{Ab initio} CC calculations using this interaction accurately predicted charge radii and binding energies of nuclei up to $^{40}$Ca and yielded a saturation point of symmetric nuclear matter consistent with the empirical one~\cite{PhysRevC.91.051301}. Further, the neutron skin in $^{48}$Ca~\cite{Hagen:2015yea} and the structure of the doubly magic nuclei $^{78}$Ni were predicted~\cite{Hagen:2016uwj}. These calculations demonstrated the significant influence of many-body observables in the construction of chiral interactions to improve the \textit{ab initio} description of the ground state properties of nuclei with the expenses of a degraded reproduction of phase shifts that results in less
accuracy in the spectroscopy of light and medium-mass nuclei. In Ref.~\cite{HUTHER2020135651}, using chiral interactions of Ref.~\cite{Entem:2017} and constraining the LECs of the 3NFs to binding energies of $^{16}$O and $^{3}$H  a new family of chiral interactions up to N${}^3$LO was defined. Using these interactions in many-body approaches, NCSM, in-medium similarity renormalization group (IM-SRG), and in-medium no-core shell model (IM-NCSM) they calculated the ground state energies and charge radii of ${}^{4}$He, ${}^{14,16,18,20,22,24,26}$O, ${}^{36,40,48,52}$Ca and ${}^{48,56,68,78}$Ni and the  spectrum of ${}^{7}$Li, ${}^{8}$Be, ${}^{9}$Be and ${}^{10}$B. The simultaneous fitting procedure was also implemented in the construction of chiral interactions up to N${}^2$LO with explicit $\Delta$-isobar included, called $\Delta$N${}^2$LO$_{\rm GO}$~\cite{Jiang:2020the}. In this work, only the LECs of the short-range 2NFs and 3NFs were determined by a simultaneous fit to the $NN$ scattering data, the properties of light nuclei up to $A = 4$  and nuclear matter. Using this chiral interaction in the CC method they calculated properties of nuclei  from $A = 16$ to $A = 132$, the equation of state for nuclear matter, and excited states of some selected nuclei. However, reproducing the unusually large charge radii of neutron-rich isotopes remained unsolved with this interaction. For the $\Delta$N${}^2$LO$_{\rm GO}$ interactions, Ref.~\cite{Nosyk:2021pxb} identified deficiencies in the $NN$ description, particularly in the P waves. 

Since about 2018 the landscape of Table~\ref{tab:lightnuclei} has broadened considerably, see Table~\ref{tab:lightnuclei2} and Refs.~\cite{Hebeler:2020ocj,Tews:2020hgp,Ekstrom:2022yea,Machleidt:2024bwl} for reviews. In the quantum Monte Carlo community, the Norfolk NV2+3 interactions of Ref.~\cite{Piarulli:2017dwd} were supplemented by the NV2+3$^{*}$ models, in which $c_D$ and $c_E$ are fixed by the binding energies of ${}^3$H and ${}^3$He and the tritium Gamow--Teller matrix element~\cite{Baroni:2018fdn}, while the local N${}^2$LO interactions of Refs.~\cite{Gezerlis:2013ipa,Lynn:2015jua,Lynn:2017fxg}, with the 3NF LECs fitted to the ${}^4$He binding energy and the $n$--${}^4$He P-wave phase shifts, were used in AFDMC calculations of nuclei up to ${}^{16}$O~\cite{Lonardoni:2017hgs,Lonardoni:2018nob} and of nuclear and neutron-star matter~\cite{Lonardoni:2019ypg}. An important finding of these studies is that the choice of the 3N contact operator and of the coordinate-space cutoff has a sizeable impact on the binding of medium-mass nuclei and on the EoS~\cite{Lonardoni:2018nob,Lonardoni:2019ypg}, showing the regulator sensitivity of 3NF effects found with non-local regulators in medium-mass nuclei~\cite{Hoppe:2019uyw,HUTHER2020135651,Hebeler:2020ocj}. Within the Darmstadt program, the strategy of Refs.~\cite{Bogner:2009un,Hebeler:2010xb} was extended to 2NFs and 3NFs at N${}^3$LO, with $c_D$ and $c_E$ fitted to the triton binding energy and to the empirical saturation point of symmetric nuclear matter~\cite{Drischler:2017wtt,Hoppe:2019uyw}, and the 1.8/2.0~(EM) interaction~\cite{Hebeler:2010xb} was used for the prediction of the drip lines for nearly 700 isotopes from He to Fe~\cite{Stroberg:2019bch}, converged calculations of ${}^{132}$Sn and ${}^{208}$Pb~\cite{Miyagi:2021pdc,Hebeler:2022aui}. Also, with the shorter-range 3N couplings constrained by the ${}^{3}$H binding energy and the ${}^{16}$O binding energy and charge radius, neutron skins of neutron-rich nuclei up to ${}^{208}$Pb were calculated~\cite{Arthuis:2024mnl}. Using similar binding-energy constraints, but without fitting nuclear radii, a 3NF with a hybrid local–nonlocal regulator was recently constructed for the Idaho local position-space 2NF, improving the description of nuclear binding energies and charge radii up to $^{132}$Sn~\cite{Hu:2026qlt}. In another recent study, the N$^3$LO$_\text{Texas}$ interaction was optimized using NN scattering data and selected few- and many-body observables, with additional oxygen and calcium binding energies used to select the final interaction. It provides a good description of nuclear binding energies, charge radii, and nuclear-matter saturation properties, while predicting the calcium neutron drip line to extend to $^{71}$Ca~\cite{Hu:2025cjl}. In the three-nucleon sector, the LENPIC collaboration established the minimum of the elastic $Nd$ differential cross section at $E_N = 70$~MeV as the most informative scattering constraint on $c_D$~\cite{LENPIC:2018ewt,Epelbaum:2019zqc,Maris:2020qne,LENPIC:2022cyu}. Studies of subleading short-range 3N interactions found substantial improvements in low-energy $pd$ polarization observables in hybrid calculations~\cite{Girlanda:2018xrw}. Related work examined 3N contact contributions induced by unitary transformations of the N${}^3$LO $NN$ interaction~\cite{Girlanda:2023znc}. As for the 3NF operators themselves, the gradient-flow regularization of Refs.~\cite{Krebs:2023ljo,Krebs:2023gge} provides a framework for deriving consistently regularized 3NFs beyond N${}^2$LO, the $D$-like 3NF at N${}^4$LO was shown to depend on 16 LECs and can be approximated using four under the assumption of $\Delta$ dominance~\cite{Huesmann:2026khj}, and the recently proposed ``new class'' of 3NFs with enhanced two-pion--two-nucleon contact couplings~\cite{Cirigliano:2024ocg} was argued to give only small contributions to the EoS of nuclear matter once scheme-dependent short-distance components are removed~\cite{Epelbaum:2026brf}.

A second major development is the transition from point estimates of $c_D$ and $c_E$ to statistically rigorous determinations with quantified uncertainties, made possible by eigenvector-continuation and related reduced-basis emulators~\cite{Ekstrom:2022yea}. Wesolowski \textit{et al.}~\cite{Wesolowski:2021cni} obtained posterior distributions for $c_D$ and $c_E$ at N${}^2$LO from the ${}^3$H binding energy, the ${}^4$He binding energy and radius and the ${}^3$H $\beta$-decay rate, demonstrating that, in that analysis, a consistent solution exists only if the $\chi$EFT truncation error is included and that the $\beta$-decay rate is essential to break the $c_D$--$c_E$ degeneracy. These posteriors were propagated to $A=6$ nuclei~\cite{Djarv:2021hcj}. More recently, Ram\'irez Jim\'enez \textit{et al.}~\cite{Jimenez:2026xgh} found that N$^2$LO calculations with $c_D$ independently constrained by nucleon–deuteron scattering overpredict the tritium Gamow–Teller matrix element. This discrepancy may reflect the fine-tuned nature of the Gamow–Teller matrix element, pointing to sizable truncation uncertainties and potentially important higher-order two-body axial-current contributions. This also highlights the importance of using different observables to constrain 3NFs and provide independent cross-checks. Related Bayesian analyses determined $NN$ LECs through N${}^3$LO and in $\Delta$-full chiral EFT, including correlated truncation errors~\cite{Svensson:2022kkj,Svensson:2023twt}. Bayesian LEC inference has also been applied to low-resolution (SRG-evolved) Hamiltonians~\cite{Plies:2025vaq} and to local chiral interactions in light nuclei using Faddeev and AFDMC emulators~\cite{Curry:2025pna}. As a methodological development in the two-nucleon sector, AFDMC emulation was demonstrated for the deuteron~\cite{Somasundaram:2024zse}. In an earlier study, the contribution of the 3NF LECs to the uncertainty budget of $n$--$\alpha$ scattering was quantified with Gaussian-process emulators in the NCSM with continuum~\cite{Kravvaris:2020lhp}. History matching was also used to constrain $\Delta$-full chiral interactions at N${}^2$LO using $NN$ scattering and selected finite-nucleus observables, leading to predictions for the neutron skin and dipole polarizability of ${}^{208}$Pb and the symmetry energy of nuclear matter~\cite{Hu:2021trw}. A related study used nuclear-matter emulators to examine the correlations between few-nucleon constraints and saturation properties~\cite{Jiang:2022oba}. The measured energies of ${}^{27,28}$O provided further tests of these interactions~\cite{Kondo:2023lty}. Ref.~\cite{Sun:2024iht} used rotational states of ${}^{24}$Ne to constrain predictions for rotational observables in other neon isotopes. The same work studied shape coexistence in ${}^{30}$Ne and ${}^{32}$Mg using the 1.8/2.0~(EM) interaction.

As noted above, {\it ab initio} nuclear many-body calculations using chiral effective field theory have made significant progress in many different nuclear structure applications in different regimes of nuclear masses. However, simultaneously describing the nucleon-nucleon phase shifts as well as the bulk properties of both nuclei and nuclear matter still remains unsolved. The results from \textit{ab initio} calculations summarized above show that the detailed features of the short-distance regulator often have a strong impact on the final results. The \textit{ab initio} calculations in Ref.~\cite{Elhatisari:2016owd,Kanada-Enyo:2020zzf} revealed that the range and degrees of locality of the nuclear interactions are the key to form the building blocks of matter. Also, regardless of the computational method or choice of chiral interaction, studies in {\it ab initio} nuclear structure theory have independently found that the properties of medium-mass nuclei and nuclear matter depend on the choice of regulator used for the three-nucleon interactions. Recently, in the framework of nuclear lattice effective field theory (NLEFT) new insights on the nuclear interactions were obtained that have the potential to solve the long-standing challenges in simultaneously reproducing an accurate description of NN scattering phase shifts, nuclear binding energies, charge radii, the equation of state of pure neutron matter, and saturation properties of symmetric nuclear matter~\cite{Elhatisari:2022zrb}. In this work, instead of regulating the three-nucleon contact interaction associated with the $c_{E}$ term using a well-defined functional form, this interaction is decomposed in Cartesian coordinates into different components associated with regulator-dependent coefficients which are determined by constraining to the nuclear binding energies. Similarly, the contact interaction associated with the $c_{D}$ term of the three-nucleon one-pion exchange potential are also decomposed into components with regulator-dependent coefficients. At the end of the procedure, the overall structure of these 3NFs can be viewed as the familiar $c_{E}$ and $c_{D}$ terms that appear at N${}^2$LO with a complicated form of the regulator. In Ref.~\cite{Elhatisari:2022zrb}, a subset of the decomposed 3NFs in Cartesian coordinates was considered, and it was found that the improvement starts to plateau after including six additional components of these considered 3NFs. In Ref.~\cite{Elhatisari:2025fyu}, the same Hamiltonian was used for the first lattice calculation of neutron--alpha scattering, and it was found that the $^2P_{1/2}$ phase shift cannot be reproduced with the 3NFs of Ref.~\cite{Elhatisari:2022zrb}, which motivates the inclusion of $n$--$\alpha$ scattering data in the determination of the 3NF LECs. In this paper, we consider the set of decomposed 3NFs in Cartesian coordinates specified in Appendix~\ref{app:cartesian} and study the dependence of the properties of light and medium-mass nuclei as well as of neutron--alpha scattering on the structure and choice of regulator of the 3NFs. In addition, we introduce a complementary short-range construction in which the interaction of a nucleon pair is projected onto definite two-nucleon partial waves and is multiplied by a scalar-isoscalar density of the third nucleon. This provides a partial-wave-resolved probe of short-distance three-body physics while avoiding a full three-body partial-wave decomposition in both Jacobi coordinates.

This paper is organized as follows. In Sec.~\ref{sec:2N}, we display the lattice Hamiltonian underlying 
our study and discuss briefly the pertinent two-nucleon forces. Sec.~\ref{sec:3N} contains a detailed
discussion of the locally and non-locally smeared three-nucleon forces, which are the central object of this work, and introduces the partial-wave-resolved three-nucleon contact interactions. In Sec.~\ref{sec:results-and-discussion}, we describe the optimization strategy and analyze the resulting nuclear binding energies and neutron--alpha scattering phase shifts, and we summarize our findings in Sec.~\ref{sec:summary}.

\section{Two nucleon forces on the lattice}
\label{sec:2N}

In this section, we summarize the two-nucleon lattice Hamiltonian employed in the present work. Our aim is to define the N${}^3$LO two-nucleon interaction that provides the fixed two-body background for the analysis of the three-nucleon sector discussed in the following sections.

In the many-nucleon calculations, the leading-order Hamiltonian is treated non-perturbatively, while higher-order contributions are evaluated in first-order perturbation theory. To test the stability of this treatment with respect to the two-nucleon reference energy, we use six reference variants of the leading-order Hamiltonian. For each observable, after adding the three-nucleon contribution we select the variant with the lowest $E+\sigma_{\rm MC}$, and the spread among the six variants is included in the quoted squared Monte Carlo uncertainty, as specified in Sec.~\ref{subsec:protocol}. This probes dependence on the reference state, and it does not establish a global variational minimum of the full Hamiltonian.

The Hamiltonian combines the two-nucleon interaction through N${}^3$LO with the leading three-nucleon interaction,
\begin{equation}
H = K 
+ V_{\rm OPE}^{\Lambda_{\pi}} 
+ V_{\rm C_{\pi}}^{\Lambda_{\pi}} 
+ V_{\rm C}
+ V_{\rm 3N}^{\rm Q^3}
+ V_{\rm 2N}^{\rm Q^4}
+ W_{\rm 2N}^{\rm Q^4}
+ V_{\rm 2N,WFM}^{\rm Q^4}
+ W_{\rm 2N,WFM}^{\rm Q^4}\,,
\label{eq:H-N${}^3$LO}
\end{equation}
where $K$ is the kinetic-energy term with nucleon mass $m=\Fixed{Mass}$ MeV, $V_{\rm OPE}^{\Lambda_{\pi}}$ and $V_{\rm C_{\pi}}^{\Lambda_{\pi}}$ represent the one-pion-exchange potential (OPEP) together with the recently developed regularization scheme of Ref.~\cite{Reinert:2017usi},
\begin{align}
V_{\rm OPE}^{\Lambda_{\pi}}  = - \frac{g_A^2}{8F^2_{\pi}}
\sum_{\vec{n}',\vec{n},S',S,I}
:\rho_{S',I}(\vec{n}')\,f_{S',S}(\vec{n}'-\vec{n}) 
\rho_{S,I}(\vec{n}): \,,
\label{eq:OPEP-full}
\end{align}
\begin{align}
V_{\rm C_{\pi}}^{\Lambda_{\pi}}  = -C_{\pi} \, \frac{g_A^2}{8F^2_{\pi}} 
\sum_{\vec{n}',\vec{n},S,I}
:\rho_{S,I}(\vec{n}')
f^{\pi}(\vec{n}'-\vec{n})
\rho_{S,I}(\vec{n}):\,.
\label{eq:OPEP-counter}
\end{align}
Here, $f^{\pi}$ is a local regulator with cutoff $\Lambda_{\pi}$ in momentum space,
\begin{align}
f^{\pi}(\vec{n}'-\vec{n})
=
\frac{1}{L^3}
\sum_{\vec{q}}
e^{-i\vec{q}\cdot(\vec{n}'-\vec{n})-(\vec{q}^2+M^2_{\pi})/\Lambda_{\pi}^2}\,,
\end{align}
and $f_{S',S}$ is the locally regulated pion correlation function,
\begin{align}
f_{S',S}(\vec{n}'-\vec{n}) 
= \frac{1}{L^3}\sum_{\vec{q}}
\frac{q_{S'}q_{S}\, e^{-i\vec{q}\cdot(\vec{n}'-\vec{n})-(\vec{q}^2+M^2_{\pi})/\Lambda_{\pi}^2}}{\vec{q}^2 + M_{\pi}^2}\,.
\label{eq:f-pi-SS}
\end{align}
The coefficient $C_{\pi}$ is given by
\begin{align}
C_{\pi} = -
\frac{\Lambda_{\pi} (\Lambda_{\pi}^2-2M_{\pi}^{2}) + 2\sqrt{\pi} M_{\pi}^3\exp(M_{\pi}^2/\Lambda_{\pi}^{2}){\rm erfc}(M_{\pi}/\Lambda_{\pi})}
{3 \Lambda_{\pi}^3}\,,
\end{align}
with $g_{A}=\Fixed{Axial}$ the axial-vector coupling constant, adjusted to account for the Goldberger--Treiman discrepancy~\cite{Fettes:1998ud}, $F_{\pi}=\Fixed{Fpi}$~MeV the pion decay constant, and $M_{\pi}=\Fixed{Mpi}$~MeV the pion mass. The term in Eq.~(\ref{eq:OPEP-counter}) is a counterterm introduced to remove the short-distance admixture in the OPEP~\cite{Reinert:2017usi}. Here, we use the notation
\begin{align}
\rho(\vec{n}) = \sum_{i,j=0,1} 
a^{\dagger}_{i,j}(\vec{n}) \, a_{i,j}(\vec{n})
\,,
\label{eqn:appx--001}
\end{align}
and
\begin{align}
\rho_{S,I}(\vec{n}) = \sum_{i,j,i^{\prime},j^{\prime}=0,1} 
a^{\dagger}_{i,j}(\vec{n}) \, [\sigma_{S}]_{ii^{\prime}} \, [\tau_{I}]_{jj^{\prime}} \, a_{i^{\prime},j^{\prime}}(\vec{n})
\label{eqn:appx--005}
\end{align}
for the density operators. Here, $a$ ($a^{\dagger}$) denotes an annihilation (creation) operator with spin index $i=0,1$ (up, down) and isospin index $j=0,1$ (proton, neutron). The term $V_{\rm C}$ denotes the Coulomb interaction, $V_{\rm 3N}^{\rm Q^3}$ the three-nucleon potential, $V_{\rm 2N}^{\rm Q^4}$ the short-range two-nucleon interaction at N${}^3$LO, $W_{\rm 2N}^{\rm Q^4}$ the Galilean-invariance-restoration (GIR) interaction at N${}^3$LO~\cite{Li:2019ldq}, $V_{\rm 2N,WFM}^{\rm Q^4}$ the wave function matching (WFM) interaction defined by $H'-H$, where $H'=U^{\dagger}HU$ is the unitarily transformed Hamiltonian whose two-body wave functions at short distances are matched to those of a simple SU(4)-symmetric Hamiltonian that is easily solvable by Monte Carlo methods~\cite{Elhatisari:2022zrb}, and $W_{\rm 2N,WFM}^{\rm Q^4}$ the corresponding GIR correction to the WFM interaction.

For details of the Coulomb interaction and the short-range two-nucleon operators, we refer the reader to Ref.~\cite{Li:2018ymw}. For the complete construction of the N${}^3$LO lattice interaction employed here, we refer to Ref.~\cite{Elhatisari:2022zrb}.

\begin{figure}[htb]
\centering\includegraphics[width=1.0\textwidth]{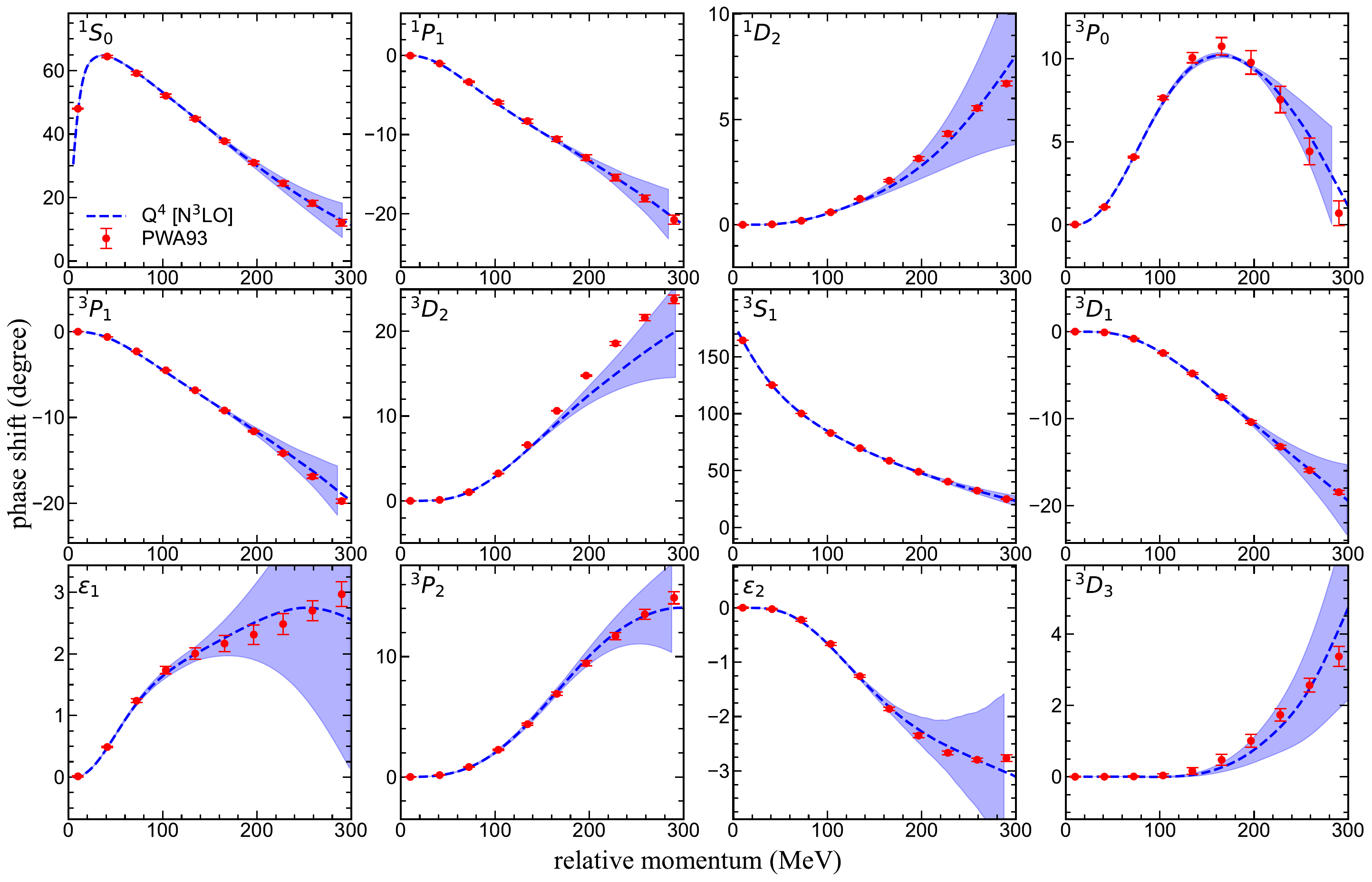}
	\caption{The  neutron-proton scattering phase shifts and mixing angles with the theoretical error bands versus the relative momentum. We use the  EKM~\cite{Epelbaum:2014efa,Epelbaum:2014sza} approach to determine the theoretical errors.}
	\label{fig:np-phase-shifts-w-errors}
\end{figure}

In the present calculations, we employ a lattice spacing of $a=\Fixed{Spacing}$~fm. The low-energy constants (LECs) of the short-range two-nucleon interaction up to N${}^3$LO are determined by reproducing the neutron--proton scattering phase shifts and mixing angles from the Nijmegen partial-wave analysis (PWA)~\cite{Stoks:1993tb}. The chosen lattice spacing corresponds to a momentum-space cutoff of about $\Fixed{Cutoff}$~MeV and defines the resolution scale at which the hidden spin-isospin symmetry of the nucleon--nucleon interaction is most accurately satisfied~\cite{Lee:2020esp}. In Fig.~\ref{fig:np-phase-shifts-w-errors}, we show the resulting neutron--proton scattering phase shifts and mixing angles up to N${}^3$LO as functions of the relative momentum, compared with the Nijmegen PWA. It is important to note that the Nijmegen PWA contains only statistical uncertainties, whereas our analysis also includes theoretical uncertainties estimated using the EKM approach~\cite{Epelbaum:2014efa,Epelbaum:2014sza}.

\section{Three nucleon forces on the lattice}
\label{sec:3N}
The three-nucleon interactions at N${}^2$LO consist of a contact potential, a one-pion exchange potential, and a two-pion exchange potential~\cite{Friar:1998zt,Epelbaum:2002vt,Epelbaum:2009zsa}, and for the lattice formulation see also Ref.~\cite{Lahde:2019npb}.

We write the two-pion-exchange three-nucleon interaction in the lattice Hamiltonian as
\begin{equation}
V_{\rm 3N}^{\rm (TPE)}
=
V_{\rm 3N}^{\rm (TPE1)}
+
V_{\rm 3N}^{\rm (TPE2)}
+
V_{\rm 3N}^{\rm (TPE3)} \, .
\end{equation}
In the present formulation, the TPE interaction is kept in the unsmeared sector and is expressed in terms of the pointlike densities given in Eqs.~(\ref{eqn:appx--001}) and (\ref{eqn:appx--005}). We then define the single-derivative pion kernel
\begin{align}
f^{\pi\pi}_{S}(\vec n'-\vec n)
=
\frac{1}{L^3}
\sum_{\vec q}
\frac{q_S\,e^{-i\vec q\cdot(\vec n'-\vec n)}\, e^{-(\vec{q}^2+M^2_{\pi})/\Lambda_{\pi}^2}}
{\vec q^{\,2}+M_\pi^2}\,,
\label{eq:f-pipi-S}
\end{align}
which corresponds to a single derivative acting on a regulated pion propagator (the notation follows Ref.~\cite{Elhatisari:2022zrb}), while the double-derivative kernel $f_{S',S}$ of Eq.~(\ref{eq:f-pi-SS}) is the same locally regulated pion correlation function that appears in the two-nucleon OPE potential. With these definitions, the lattice TPE three-nucleon interaction takes the form
\begin{align}
V_{\rm 3N}^{\rm (TPE1)}
= &
\frac{c_{3}}{F_{\pi}^{2}}
\left(\frac{g_{A}}{2F_{\pi}}\right)^{2}
\, 
\sum_{S,S^{\prime},S^{\prime\prime},I}
\sum_{\vec{n},\vec{n}^{\prime},\vec{n}^{\prime\prime}}
\nonumber \\
&
\times	
\, : \, 
\rho_{S^{\prime},I}(\vec{n}^{\prime}) \,
f_{S^{\prime},S}(\vec{n}^{\prime}-\vec{n})
f_{S^{\prime\prime},S}(\vec{n}^{\prime\prime}-\vec{n})
\rho_{S^{\prime\prime},I}(\vec{n}^{\prime\prime}) \,
\rho(\vec{n})
\, : \, \,,
\label{eqn:V_TPE1^3N--001}
\end{align}

\begin{align}
V_{\rm 3N}^{\rm (TPE2)}
= &
-\frac{2c_{1}M_{\pi}^{2}}{F_{\pi}^{2}}
\left(\frac{g_{A}}{2F_{\pi}}\right)^{2}
\, 
\sum_{S,S^{\prime},I}
\sum_{\vec{n},\vec{n}^{\prime},\vec{n}^{\prime\prime}}
\nonumber \\
&
\times	
\, : \, 
\rho_{S^{\prime},I}(\vec{n}^{\prime}) \,
f_{S^{\prime}}^{\pi\pi}(\vec{n}^{\prime}-\vec{n})
f_{S}^{\pi\pi}(\vec{n}^{\prime\prime}-\vec{n})
\rho_{S,I}(\vec{n}^{\prime\prime}) \,
\rho(\vec{n})
\, : \, 
\,,
\label{eqn:V_TPE2^3N--001}
\end{align}
\begin{align}
V_{\rm 3N}^{\rm (TPE3)}
= \frac{c_{4}}{2F_{\pi}^{2}}
&
\left( \frac{g_{A}}{2 F_{\pi}}\right)^{2}
\sum_{S_{1},S_{2},S_{3}}
\sum_{I_{1},I_{2},I_{3}}
\sum_{S^{\prime},S^{\prime\prime}}
\sum_{\vec{n},\vec{n}^{\prime},\vec{n}^{\prime\prime}}
\varepsilon_{S_1,S_2,S_3}
\varepsilon_{I_1,I_2,I_3}
\nonumber \\
&
\times	
\, : \, 
\rho_{S^{\prime},I_{1}}(\vec{n}^{\prime}) \,
f_{S^{\prime},S_{1}}(\vec{n}^{\prime}-\vec{n})
f_{S^{\prime\prime},S_{2}}(\vec{n}^{\prime\prime}-\vec{n})
\rho_{S^{\prime\prime},I_{2}}(\vec{n}^{\prime\prime}) \,
\rho_{S_{3},I_{3}}(\vec{n})
\, : \, \,.
\label{eqn:V_TPE3^3N--001}
\end{align}
The corresponding low-energy constants are fixed from pion--nucleon scattering. We use the N$^3$LO values (in the counting appropriate for nuclear forces, ``NN counting'') of the Roy--Steiner analysis of Refs.~\cite{Hoferichter:2015tha,Hoferichter:2015hva},
\begin{equation}
c_{1}=-1.10(3)\ {\rm GeV}^{-1},\qquad
c_{3}=-5.54(6)\ {\rm GeV}^{-1},\qquad
c_{4}=4.17(4)\ {\rm GeV}^{-1},
\end{equation}
which are also employed in the N$^4$LO two-nucleon potentials of Ref.~\cite{Reinert:2017usi}. Using instead the N$^2$LO values of the same analysis would shift the TPE contribution by an amount that is absorbed in the refit of the short-range three-nucleon LECs discussed below. In the present work we take $\Lambda_\pi=\Fixed{PionCutoff}~{\rm MeV}$.

Now we introduce the general forms of the purely local (in terms of the locally smeared densities $\rho^{(d)}$ defined in Eqs.~(\ref{eqn:rho-smeared-001}) and (\ref{eqn:rho-smeared-005}) below),
\begin{align}
[V_{c_{E}}^{(d)}]_{abc} = [c_{E}^{(d)}]_{abc} 
\, \sum_{\vec{n},\vec{n}^{\prime},\vec{n}^{\prime\prime}}
\, : \, \rho^{(d)}(\vec{n})\,
\rho^{(d)}(\vec{n}^{\prime}) \, 
\rho^{(d)}(\vec{n}^{\prime\prime}) \, : \, \delta_{|\vec{n}-\vec{n}^{\prime}|^2,a} \, \,  \delta_{|\vec{n}-\vec{n}^{\prime\prime}|^2,b} \, \,  \delta_{|\vec{n}^{\prime}-\vec{n}^{\prime\prime}|^2,c},  
\label{eqn:V_cE-sL-001}      
\end{align}
\begin{align}
V_{c_{D}}^{(d)}
= -\frac{c_{D}^{(d)} \, g_{A}}{4F_{\pi}^{4} \Lambda_{\chi}}\,  \sum_{\vec{n},S,I}
\sum_{\vec{n}^{\prime},S^{\prime}}
\, : \,
\rho^{(0)}_{S^{\prime},I}(\vec{n}^{\prime})
f_{S^{\prime},S}(\vec{n}^{\prime}-\vec{n})
\rho^{(d)}_{S,I}(\vec{n})
\rho^{(d)}(\vec{n})
\, : \,
\,,
\label{eqn:V_cD-sL-001} 
\end{align}
and local and nonlocal,
\begin{align}
[V_{c_{E}}^{(d)}]^{s_{\rm NL}}_{abc} = [c_{E}^{(d)}]^{s_{\rm NL}}_{abc} 
\, \sum_{\vec{n},\vec{n}^{\prime},\vec{n}^{\prime\prime}}
\, : \, \tilde{\rho}^{(d)}(\vec{n})\,
\tilde{\rho}^{(d)}(\vec{n}^{\prime}) \, 
\tilde{\rho}^{(d)}(\vec{n}^{\prime\prime}) \, : \, \delta_{|\vec{n}-\vec{n}^{\prime}|^2,a} \, \,  \delta_{|\vec{n}-\vec{n}^{\prime\prime}|^2,b} \, \,  \delta_{|\vec{n}^{\prime}-\vec{n}^{\prime\prime}|^2,c},  
\label{eqn:V_cE-sNL-001} 
\end{align}
\begin{align}
V_{c_{D},s_{\rm NL}}^{(d)}
= -\frac{c_{D,s_{\rm NL}}^{(d)} \, g_{A}}{4F_{\pi}^{4} \Lambda_{\chi}}\,  \sum_{\vec{n},S,I}
\sum_{\vec{n}^{\prime},S^{\prime}}
\, : \,
\rho^{(0)}_{S^{\prime},I}(\vec{n}^{\prime})
f_{S^{\prime},S}(\vec{n}^{\prime}-\vec{n})
\tilde{\rho}^{(d)}_{S,I}(\vec{n})
\tilde{\rho}^{(d)}(\vec{n})
\, : \,
\,,
\label{eqn:V_cD-sNL-001} 
\end{align}
three-nucleon contact and three-nucleon one-pion exchange potentials which can be decomposed in Cartesian coordinates into their weighted components by the regulator-dependent coefficients. Here, $\Lambda_\chi = \Fixed{ChiralScale}$~MeV is the chiral symmetry breaking scale, $\Lambda_\chi \simeq M_\rho$, as in Refs.~\cite{Epelbaum:2002vt,Epelbaum:2009zsa}. Note that with the normalization of Eqs.~(\ref{eqn:V_cD-sL-001}) and (\ref{eqn:V_cD-sNL-001}), which follows Ref.~\cite{Elhatisari:2022zrb}, the coefficient $c_D^{(d)}$ corresponds to $c_D/2$ in the standard normalization of Refs.~\cite{Epelbaum:2002vt,Epelbaum:2009zsa}, $V_D=-\frac{g_A}{8F_\pi^2}\frac{c_D}{F_\pi^2\Lambda_\chi}\sum_{i\neq j\neq k}\frac{(\vec\sigma_j\cdot\vec q_j)(\vec\sigma_i\cdot\vec q_j)}{\vec q_j^{\,2}+M_\pi^2}\,\boldsymbol\tau_i\cdot\boldsymbol\tau_j$.

The coefficients $[c_E^{(d)}]_{abc}$ carry the dimension of energy, and the normalization used when quoting their fitted values is specified in Sec.~\ref{sec:results-and-discussion}. For the pointlike on-site term, the Pauli principle restricts the three nucleons to the totally antisymmetric spin--isospin state, on which all six possible spin--isospin structures of a three-nucleon contact interaction are proportional to each other~\cite{Epelbaum:2002vt,Epelbaum:2009zsa,Girlanda:2011fh}. The SU(4)-symmetric form of Eq.~(\ref{eqn:V_cE-sL-001}) is therefore equivalent to the conventional form $V_E=\frac{c_E}{2F_\pi^4\Lambda_\chi}\sum_{j\neq k}\boldsymbol\tau_j\cdot\boldsymbol\tau_k$ of Ref.~\cite{Epelbaum:2002vt}, with $[c_E^{(0)}]_{000}=-c_E/(2F_\pi^4\Lambda_\chi)$, and in the normalization of Ref.~\cite{Elhatisari:2022zrb}, $V_{c_E}^{(d)}=\frac{1}{6}\frac{c_E^{(d)}}{2F_\pi^4\Lambda_\chi}\sum_{\vec n}:[\rho^{(d)}(\vec n)]^3:$, one has correspondingly $[c_E^{(d)}]_{000}=c_E^{(d)}/(12F_\pi^4\Lambda_\chi)$. For the smeared ($d>0$, $s_{\rm NL}>0$) and spatially extended ($abc\neq000$) configurations introduced below and in Appendix~\ref{app:cartesian} this equivalence no longer holds, and the SU(4)-symmetric form is a definite choice, motivated by the approximate Wigner SU(4) symmetry of the short-range nuclear interaction~\cite{Lee:2020esp}. 
In Eqs.~(\ref{eqn:V_cE-sL-001}) and (\ref{eqn:V_cE-sNL-001}), the indices $a$, $b$ and $c$ denote the square of the absolute relative distances between three-nucleon configurations. The $\tilde{\rho}$'s are density operators that are smeared both locally and non-locally,
\begin{equation}
\tilde{\rho}^{(d)}(\vec{n}) = \sum_{i,j=0,1} 
\tilde{a}^{\dagger}_{i,j}(\vec{n}) \, \tilde{a}^{\,}_{i,j}(\vec{n})
+
s_{\rm L}
\sum_{1\le |\vec{n}-\vec{n}^{\prime}|^2 \le d} 
\,
\sum_{i,j=0,1} 
\tilde{a}^{\dagger}_{i,j}(\vec{n}^{\prime}) \, \tilde{a}^{\,}_{i,j}(\vec{n}^{\prime})
\,,
\end{equation}
\begin{align}
\tilde{\rho}^{(d)}_{S,I}(\vec{n}) = & \sum_{i,j,i^{\prime},j^{\prime}=0,1} 
\tilde{a}^{\dagger}_{i,j}(\vec{n}) \, [\sigma_{S}]_{ii^{\prime}} \, [\tau_{I}]_{jj^{\prime}} \, \tilde{a}^{\,}_{i^{\prime},j^{\prime}}(\vec{n})
\nonumber \\
& +
s_{\rm L}
\sum_{1\le |\vec{n}-\vec{n}^{\prime}|^2 \le d} 
\,
\sum_{i,j,i^{\prime},j^{\prime}=0,1} 
\tilde{a}^{\dagger}_{i,j}(\vec{n}^{\prime}) \, [\sigma_{S}]_{ii^{\prime}} \, [\tau_{I}]_{jj^{\prime}} \, \tilde{a}^{\,}_{i^{\prime},j^{\prime}}(\vec{n}^{\prime}) \,,
\end{align} 
where the nonlocally smeared annihilation and creation operators with the nonlocal smearing parameter $s_{\rm NL}$ are given as,
\begin{align}
\tilde{a}_{i,j}(\vec{n})=a_{i,j}(\vec{n})+s_{\rm NL}\sum_{|\vec{n}^{\prime}-\vec{n}|=1}a_{i,j}(\vec{n}^{\prime})\,,\\
\tilde{a}^{\dagger}_{i,j}(\vec{n})=a^{\dagger}_{i,j}(\vec{n})+s_{\rm NL}\sum_{|\vec{n}^{\prime}-\vec{n}|=1}a^{\dagger}_{i,j}(\vec{n}^{\prime})\,.
\end{align}
Also the locally smeared density operators are defined as, 
\begin{align}
\rho^{(d)}(\vec{n}) = \sum_{i,j=0,1} 
a^{\dagger}_{i,j}(\vec{n}) \, a^{\,}_{i,j}(\vec{n})
+
s_{\rm L}
\sum_{1\le |\vec{n}-\vec{n}^{\prime}|^2 \le d} 
\,
\sum_{i,j=0,1} 
a^{\dagger}_{i,j}(\vec{n}^{\prime}) \, a^{\,}_{i,j}(\vec{n}^{\prime})
\label{eqn:rho-smeared-001}
\end{align}
\begin{align}
\rho^{(d)}_{S,I}(\vec{n}) = & \sum_{i,j,i^{\prime},j^{\prime}=0,1} 
a^{\dagger}_{i,j}(\vec{n}) \, [\sigma_{S}]_{ii^{\prime}} \, [\tau_{I}]_{jj^{\prime}} \, a^{\,}_{i^{\prime},j^{\prime}}(\vec{n})
\nonumber \\
& +
s_{\rm L}
\sum_{1\le |\vec{n}-\vec{n}^{\prime}|^2 \le d} 
\,
\sum_{i,j,i^{\prime},j^{\prime}=0,1} 
a^{\dagger}_{i,j}(\vec{n}^{\prime}) \, [\sigma_{S}]_{ii^{\prime}} \, [\tau_{I}]_{jj^{\prime}} \, a^{\,}_{i^{\prime},j^{\prime}}(\vec{n}^{\prime}) \,,
\label{eqn:rho-smeared-005}
\end{align} 
where the superscript $d$ specifies the range of the local smearing, so that $\rho^{(0)}(\vec n)=\rho(\vec n)$ and $\rho^{(0)}_{S,I}(\vec n)=\rho_{S,I}(\vec n)$ coincide with the pointlike densities of Eqs.~(\ref{eqn:appx--001}) and (\ref{eqn:appx--005}).

\begin{table}[ht]
	\caption{The locally smeared three-nucleon one-pion exchange potential. The pion exchange is depicted by a blue dashed line.}
	\label{tab:cD-sL-config}
	\centering
	\begin{tabular}{ |c|| c | c| c | c| }
		\hline\hline		
		Coefficients  
		&    \csuse{C_cD_sL0}   & \csuse{C_cD_sL1} 	& \csuse{C_cD_sL2}  &  \csuse{C_cD_sL3}   \\\hline
		$s_{{\rm L}}$  
		&   \Fixed{LocalSmearing}   & \Fixed{LocalSmearing} 	& \Fixed{LocalSmearing} & \Fixed{LocalSmearing}     \\
		$R_{s_{{\rm L}}}$~(fm)  
		&   0.00   & \Fixed{Spacing} 	& \Fixed{LocalRange2} & \Fixed{LocalRange3}  \\\hline
		\makecell{Shape} &  \makecell{\csuse{V_cD_sL0pic}}
		&    \makecell{\csuse{V_cD_sL1pic}}
		&   \makecell{\csuse{V_cD_sL2pic}} 
		&  \makecell{\csuse{V_cD_sL3pic}} \\\hline\hline
	\end{tabular}
\end{table}
In Eqs.~(\ref{eqn:V_cD-sL-001}) and (\ref{eqn:V_cD-sNL-001}) we define the three-nucleon one-pion exchange potential by regulating the contact term associated with $c_{D}$ using local smearing and nonlocal smearing, respectively, in configuration space. We decompose Eq.~(\ref{eqn:V_cD-sL-001}) for $d = 3$, the locally smeared up to next-to-next-to-nearest neighboring distance, into four components  with the smearing strength $s_{\rm L}$ and with the range of the smearing $R_{s_{\rm L}}$ as illustrated in Tab.~\ref{tab:cD-sL-config}. The first one has a purely contact interaction with no smearing and coefficient~\csuse{C_cD_sL0}, the second one is locally smeared up to nearest-neighbor lattice site and coefficient~\csuse{C_cD_sL1}, the third one is locally smeared up to next-to-nearest-neighbor lattice site and coefficient~\csuse{C_cD_sL2}, and the fourth one is locally smeared up to next-to-next-to-nearest-neighbor lattice site and coefficient~\csuse{C_cD_sL3}.
\begin{table}[ht]
	\caption{The non-locally smeared three-nucleon one-pion exchange potential. See Tab.~\ref{tab:cD-sL-config} for further notation.}
	\label{tab:cD-sNL-config}
	\centering
	\begin{tabular}{ |c|| c | c| c| }
		\hline\hline				
		Coefficients  
		& \csuse{C_cD_sNL01} 	& \csuse{C_cD_sNL02} & \csuse{C_cD_sNL03}  \\\hline
		$s_{{\rm NL}}$
		& 0.1 	& 0.2 & 0.3 \\
		$R_{s_{{\rm NL}}}$~(fm)  
		& \Fixed{Spacing} 	& \Fixed{Spacing} & \Fixed{Spacing}\\\hline
		\makecell{Shape}	&
		\makecell{\csuse{V_cD_sNL01pic}}   &
		\makecell{\csuse{V_cD_sNL02pic}}   & 
		 \makecell{\csuse{V_cD_sNL03pic}} \\\hline\hline
	\end{tabular}
\end{table}
For the nonlocal version of this potential, from  Eq.~(\ref{eqn:V_cD-sNL-001}) with $d = 0$ ($R_{s_{\rm L}}=0$) we define three different forms that are nonlocally smeared up to only nearest neighboring distance with different smearing strength and with the range of the smearing $R_{s_{\rm NL}}$ as illustrated in Tab.~\ref{tab:cD-sNL-config}. In these three potentials, the creation and annihilation operators of particles participating to the contact term are nonlocally smeared with $s_{\rm NL} = s_{\rm NL1}=0.1$, $s_{\rm NL} = s_{\rm NL2}=0.2$ and $s_{\rm NL} = s_{\rm NL3}=0.3$, respectively, and the coefficients are \csuse{C_cD_sNL01}, \csuse{C_cD_sNL02} and \csuse{C_cD_sNL03}, in order. The dashed-lines shown in the illustrative figures in Tabs.~\ref{tab:cD-sL-config} and~\ref{tab:cD-sNL-config} are just to depict long-range one-pion exchange linked to the short-range interaction of two nucleons.

Now we define the three-nucleon contact potential given in Eqs.~(\ref{eqn:V_cE-sL-001}) and (\ref{eqn:V_cE-sNL-001}). We start with the decomposition of Eq.~(\ref{eqn:V_cE-sL-001}) for $a = b = c =0$ and $d = 3$, the locally smeared three-nucleon contact potential up to next-to-next-to-nearest neighboring
distance, into four components  with the smearing strength $s_{\rm L}$ and with the range of the smearing $R_{s_{\rm L}}$ as illustrated in Tab.~\ref{tab:cE-sL-config}. 
\begin{table}[ht]
	\caption{The locally smeared  three-nucleon contact potential.}
	\label{tab:cE-sL-config}
	\centering
	\begin{tabular}{ |c|| c | c| c | c| }
		\hline\hline		
		Coefficients  
		&    \csuse{C_cE_sL0}   & \csuse{C_cE_sL1} 	& \csuse{C_cE_sL2}  &  \csuse{C_cE_sL3}   \\\hline
		$s_{{\rm L}}$  
		&   \Fixed{LocalSmearing}   & \Fixed{LocalSmearing} 	& \Fixed{LocalSmearing} & \Fixed{LocalSmearing}     \\
		$R_{s_{{\rm L}}}$~(fm)  
		&   0.00   & \Fixed{Spacing} 	& \Fixed{LocalRange2} & \Fixed{LocalRange3}  \\\hline
		\makecell{Shape} &  \makecell{\csuse{V_cE_sL0pic}}
		&    \makecell{\csuse{V_cE_sL1pic}}
		&   \makecell{\csuse{V_cE_sL2pic}} 
		&  \makecell{\csuse{V_cE_sL3pic}} \\\hline\hline
	\end{tabular}
\end{table}
These are a purely contact interaction with no smearing and coefficient~\csuse{C_cE_sL0}, a locally smeared up to nearest-neighbor lattice site and with coefficient~\csuse{C_cE_sL1}, a locally smeared up to next-to-nearest-neighbor lattice site and  with coefficient~\csuse{C_cE_sL2}, and locally smeared up to next-to-next-to-nearest-neighbor lattice site and  with coefficient~\csuse{C_cE_sL3}.
\begin{table}[ht]
	\caption{The non-locally smeared three-nucleon contact potential.}
	\label{tab:cE-sNL-config}
	\centering
	\begin{tabular}{ |c|| c | c| c| }
		\hline\hline				
		Coefficients  
		& \csuse{C_cE_sNL01} 	& \csuse{C_cE_sNL02} & \csuse{C_cE_sNL03} \\\hline
		$s_{{\rm NL}}$
		& 0.1 	& 0.2 & 0.3 \\
		$R_{s_{{\rm NL}}}$~(fm)  
		& \Fixed{Spacing} 	& \Fixed{Spacing} & \Fixed{Spacing}\\\hline
		\makecell{Shape}	&
		\makecell{\csuse{V_cE_sNL01pic}}   &
		\makecell{\csuse{V_cE_sNL02pic}}   &   
		 \makecell{\csuse{V_cE_sNL03pic}} \\\hline\hline
	\end{tabular}
\end{table}
For the nonlocally smeared forms of the three-nucleon contact potential, from  Eq.~(\ref{eqn:V_cE-sNL-001}) with $d = 0$ ($R_{s_{\rm L}}=0$) we define three different potentials that are smeared up to only nearest neighboring distance with different smearing strength and with the range of the smearing $R_{s_{\rm NL}}$ as illustrated in Tab.~\ref{tab:cE-sNL-config}.

The three-nucleon contact interaction of Eq.~(\ref{eqn:V_cE-sL-001}) can furthermore be decomposed in Cartesian coordinates, i.e.\ according to the geometry of the three nucleons ($abc\neq000$), into thirteen SU(4)-symmetric configurations that fit into a cube of side length two lattice units, supplemented by locally smeared versions of the four configurations in which two nucleons share a lattice site. Their definitions, illustrations and root-mean-square radii are collected in Appendix~\ref{app:cartesian}. Two of these configurations were part of the three-nucleon interaction of Ref.~\cite{Elhatisari:2022zrb} and the configurations considered here were employed in Ref.~\cite{Ren:2025vpe}. We include these finite-range lattice structures in the primary fit and compare with the set of 23 terms without them in Appendix~\ref{app:no-cartesian-results}.

\subsection{Partial-wave-resolved three-nucleon contact interactions}
\label{subsec:PW3NF}

The smeared $c_D$ and $c_E$ operators of Tabs.~\ref{tab:cD-sL-config}--\ref{tab:cE-sNL-config} resolve the interaction through their spatial regulators, and the $c_E$ operators and the Cartesian contact configurations of Appendix~\ref{app:cartesian} are SU(4) symmetric. As a complementary probe, we introduce {\em partial-wave-resolved three-nucleon contact interactions}. A pair operator with definite two-nucleon partial-wave labels is coupled to the scalar--isoscalar density $\rho^{(d)}(\vec n)$ of the third nucleon. This construction requires no full three-body partial-wave decomposition in both Jacobi coordinates~\cite{Hebeler:2020ocj}.

Let $A^{(n)}_{LSJM;II_z}(\vec n)$ denote a regulated pair-annihilation operator with orbital angular momentum $L$, spin $S$, total angular momentum $J$ and isospin $I$, constructed from nonlocally smeared nucleon fields by coupling the spin and orbital projections with Clebsch--Gordan coefficients as in Ref.~\cite{Li:2018ymw}. The non-negative integer $n$ counts Laplacian insertions, and $\vec n$ is the reference site of the lattice pair operator. In the continuum limit and pair centre-of-mass frame its orbital momentum dependence is proportional to $p^{L+2n}Y_{Lm}(\hat p)$ times a short-distance regulator. We consider the physical pair channels satisfying $(-1)^{L+S+I}=-1$. The partial-wave labels refer to the continuum channels, and rotational symmetry on the lattice is reduced to the cubic group. A pair operator is specified by
\begin{equation}
\alpha=\{I,S,J,L_i,L_f,n_i,n_f\}\,,
\label{eq:PW-alpha}
\end{equation}
and for each $\alpha$ we define the Hermitian partial-wave contact operator
\begin{align}
{\cal O}^{\rm PW}_{\alpha}(\vec n)
=\frac{1}{2}\sum_{M,I_z}
\bigg[
A^{(n_f)\dagger}_{L_fSJM;II_z}(\vec n)
A^{(n_i)}_{L_iSJM;II_z}(\vec n)
+
A^{(n_i)\dagger}_{L_iSJM;II_z}(\vec n)
A^{(n_f)}_{L_fSJM;II_z}(\vec n)
\bigg]\,,
\label{eq:PW-pair-contact}
\end{align}
where for $L_i\neq L_f$ (coupled channels) or $n_i\neq n_f$ the two terms are the Hermitian-conjugate transitions, the latter corresponding to the familiar $p^{2}+p'^{2}$ structure of the two-nucleon contact interactions. The three-nucleon interaction is then obtained by attaching the locally smeared scalar--isoscalar density of a third nucleon to the reference site of the pair,
\begin{align}
V^{(d)}_{{\rm 3N},\alpha}
=
D^{(d)}_{\alpha}
\sum_{\vec n}
:\!{\cal O}^{\rm PW}_{\alpha}(\vec n)\,
\rho^{(d)}(\vec n)\!:\, ,
\label{eq:PW-density-3NF}
\end{align}
where the LEC $D^{(d)}_{\alpha}$ has the dimension of an energy in the same units as the $[c_E^{(d)}]_{abc}$. The normal ordering ensures that Eq.~(\ref{eq:PW-density-3NF}) is a three-nucleon interaction, and since $\rho^{(d)}$ is summed over all spin and isospin states, the third nucleon enters as an SU(4) scalar while all dependence on $I$, $S$, $J$ and the orbital angular momenta resides in the pair. The operator is Hermitian, translation invariant, and conserves isospin and parity. Its continuum angular-momentum labels do not imply exact rotational or Galilean invariance at finite lattice spacing.

For reference, a three-nucleon contact with $d_\alpha$ derivatives has the chiral index $\nu_\alpha=3+d_\alpha$ in Weinberg counting, where
the leading 3NF has $\nu=3$ (N${}^2$LO)~\cite{Weinberg:1990,Weinberg:1991,Epelbaum:2008ga}.
The terms considered here
form a restricted set of short-range lattice interactions, and they are not a complete basis interaction at any given chiral order. The listed orders classify their
derivative structures and are not imposed as suppression factors in the fit. We consider the nine operators listed in Tab.~\ref{tab:PW3NF-orders}: two
$S$-wave contacts without derivatives, two $S$-wave contacts with one
Laplacian insertion, four $P$-wave contacts, and one
${}^3S_1$--${}^3D_1$ transition. We denote the LECs of the two
derivative-free operators by $\tilde D_{^{2S+1}L_J}$ and those of the
seven derivative operators by $D_{^{2S+1}L_J}$. The common smearing index
$d$ of the third-nucleon density is suppressed in this notation.

The nine LECs are optimized together with those of the smeared operators
and the Cartesian configurations defined in Appendix~\ref{app:cartesian}.
All three-nucleon contributions are evaluated in first-order perturbation
theory. The coefficient bounds and common fitting protocol are given in
Sec.~\ref{sec:results-and-discussion}.
\begin{table}[ht]
\caption{Partial-wave-resolved three-nucleon contact interactions considered in this work. The derivative order $d_\alpha$ and the pair channels refer to the partial-wave-projected interacting pair. The last column lists the LECs $D^{(d)}_\alpha$ of Eq.~(\ref{eq:PW-density-3NF}) in the shorthand notation used in the following sections.}
\label{tab:PW3NF-orders}
\centering
\begin{tabular}{|c||c|c|c|c|}
\hline\hline
$d_{\alpha}$ & $\nu_\alpha$ (order) & Pair channels $\alpha$ & Number & LECs \\
\hline
$0$ & $3$ (N${}^2$LO) & $\PWoneSzero$, $\PWthreeSone$ & $2$ &
\csuse{D_1S0}, \csuse{D_3S1} \\\hline
$2$ & $5$ (N${}^4$LO) &
\makecell{$\PWoneSzero$, $\PWthreeSone$ with $(n_i,n_f)=(0,1)$; \\ $\PWonePone$, $\PWthreePzero$, $\PWthreePone$, $\PWthreePtwo$; \\ $\PWthreeSoneDone$ with $(L_i,L_f)=(0,2)$} & $7$ &
\makecell{\csuse{D_1S0_01}, \csuse{D_3S1_01}; \\ \csuse{D_1P1}, \csuse{D_3P0}, \csuse{D_3P1}, \csuse{D_3P2}; \\ \csuse{D_3S1_3D1}} \\
\hline\hline
\end{tabular}
\end{table}

\providecommand{\FITSCORE}{$\mathcal{R}_{\rm fit}$}
\providecommand{\dRMSD}{$\delta_{\rm RMSD}$}
\section{Results and discussion}
\label{sec:results-and-discussion}

We constrain the short-range three-nucleon interactions using nuclear
binding energies and finite-volume neutron--alpha energies. We consider
40 three-nucleon terms: seven smeared $c_D$ terms, seven smeared $c_E$
terms (Tabs.~\ref{tab:cD-sL-config}--\ref{tab:cE-sNL-config}), nine
partial-wave-resolved contact terms with up to two derivatives
(Tab.~\ref{tab:PW3NF-orders}), and the 17 spatially extended contact
configurations defined in App.~\ref{app:cartesian}. For comparison,
we also perform fits using only the first 23 terms, without the
spatially extended configurations, and present these results in
App.~\ref{app:no-cartesian-results}.

\subsection{Optimization}
\label{subsec:protocol}
For fixed reference variants, each underlying lattice energy depends linearly on the selected LECs,
\begin{align}
 E_j^{\rm th}(\vec c)=E_{j,\rm ref}+\sum_{k\in\mathcal F_p}c_k\langle O_k\rangle_j\,.
 \label{eqn:optimization-eqn-010-new}
\end{align}
Here $\mathcal F_p$ is a selected set of $p$ short-range three-nucleon terms, and $E_{j,\rm ref}$ includes the N${}^3$LO two-nucleon interaction, Coulomb interaction, and two-pion-exchange 3NF with fixed $c_{1,3,4}$. The matrix elements are first-order perturbative expectation values in the wave-function-matching framework~\cite{Elhatisari:2022zrb}. As described in Sec.~\ref{sec:2N}, each energy is evaluated for six reference variants. After adding the 3NF, the variant minimizing $E_{j,v}+\sigma_{{\rm MC},j,v}$ is selected, and its uncertainty is
\begin{align}
 \sigma_j^2=\sigma_{{\rm MC},j}^2+\frac16\sum_{v=1}^{6}(E_{j,v}-\langle E_j\rangle_v)^2\,.
 \label{eq:variant-error}
\end{align}
Here $\langle E_j\rangle_v=\frac16\sum_{v=1}^{6}E_{j,v}$,
and $\sigma_{{\rm MC},j}$ is the Monte Carlo uncertainty of the
selected variant.
The Monte Carlo variance includes the squared coefficient-weighted errors of the operator matrix elements, added in quadrature. No chiral truncation uncertainty enters this fit or the result error bars.

The fit uses $N_{\rm obs}=38$ energy observables $z_n$, which are ground-state energies per nucleon $E/A$ for 26 nuclei, and twelve relative finite-volume energies $\Delta E_L=E_L({}^5{\rm He})-E_L({}^4{\rm He})$ in the $^2P_{3/2}$ and $^2P_{1/2}$ channels, with $L=7,\ldots,12$. The nuclei are ${}^3$H, ${}^{3,4,6}$He, ${}^{6,7}$Li, ${}^{8-10}$Be, ${}^{10,11}$B, ${}^{12-14}$C, ${}^{14,15}$N, ${}^{16-18}$O, ${}^{18}$F, ${}^{20}$Ne, ${}^{24}$Mg, ${}^{28}$Si, ${}^{32}$S, ${}^{36}$Ar, and ${}^{40}$Ca. The empirical scattering energies follow from L\"uscher's condition and the $R$-matrix phase shifts~\cite{Elhatisari:2025fyu,hale2008private}, and each ${}^5$He level is paired with ${}^4$He in the same box. For a nuclear ground state, $z_n^{\rm th}=E_j^{\rm th}/A_j$
and $s_n^2=\sigma_j^2/A_j^2$, and for a neutron--alpha level,
$z_n^{\rm th}=E_{L,J}^{\rm th}({}^5{\rm He})-E_L^{\rm th}({}^4{\rm He})$
and
$s_n^2=\sigma_{L,J}^2({}^5{\rm He})+\sigma_L^2({}^4{\rm He})$,
where $J=1/2,3/2$ distinguishes the two scattering channels.
The reference values $z_n^{\rm exp}$ are defined in the same way.
Covariances between the underlying energies are omitted.
Here $n$ labels the quantities entering the fit, whereas $j$ in
Eqs.~(\ref{eqn:optimization-eqn-010-new}) and
(\ref{eq:variant-error}) labels the underlying lattice energies.

We divide the data into a training set $\mathcal S_{\rm trn}$,
used to determine the LECs, and a test set $\mathcal S_{\rm tst}$.
For each $p=2,\ldots,8$, we perform the following stochastic search.
\begin{enumerate}
\item Initially assign each observable to the training set with
probability $1/2$ and choose an acceptable set of $p$ 3NF terms.
The triton and ${}^4$He are always in the training set, and
${}^3$He is always in the test set. In the fits, the data with
$L=8$--$12$ in $^2P_{3/2}$ and $L=8$--$10$
in $^2P_{1/2}$ are always fitted, while the data with $L=11,12$ in
$^2P_{1/2}$ is not constrained due to the large statistical noise in the data. An acceptable combination contains
at least one $c_D$-type and one $c_E$-type term, and the spatially
extended contact terms count as $c_E$-type, while the partial-wave
terms satisfy neither requirement. Combinations saved as the best
fit of three earlier chains are excluded from further initial
selections and replacements of terms.
\item 
For each proposed training set and combination of 3NF terms, we determine
the LECs by bounded least-squares fitting with unit observable weights.
The coefficient bounds are
\begin{align}
 |c_k|\le c_k^{\max}
 =0.25\operatorname{median}_{n}
 \left|\frac{V_{n,\rm ref}}{X_{nk}}\right|\,.
 \label{eq:naturalness-bound}
\end{align}
Here $X_{nk}$ is the contribution of term $k$ to observable $z_n$
per unit coefficient, and $V_{n,\rm ref}$ is the reference
contribution with its kinetic energy removed, and the factor 0.25 is to ensure that the contributions of individual 3N terms do not
exceed 25\% of the total contribution from the 2N interactions. Both quantities are
divided by $A$ for nuclear ground states and formed as the
${}^5$He--${}^4$He difference for scattering levels.
\item The search uses the Gaussian likelihood
\begin{equation}
\log\mathcal L=-\frac{1}{2}
\sum_{n\in\mathcal S_{\rm trn}}
\left[
\frac{(z_n^{\rm th}-z_n^{\rm exp})^2}{s_n^2}
+\log(2\pi s_n^2)
\right],
\label{eqn:optimization-eqn-020-new}
\end{equation}
Here $s_n^2$ is the variance defined above. At each iteration, we make one to four
random attempts to switch observables between the training and test
sets. We then refit the LECs and accept the proposed
assignment with probability $\min(1,\mathcal L'/\mathcal L)$.
Every 20 iterations, we also make ten attempts to replace one selected
3NF term by another, using the same acceptance rule.
\item Run each chain for $60\,000$ iterations with $10\,000$ iterations for the thermalization. Then, we record a fit only if it improves the lowest previously recorded score
and satisfies ${\rm RMSD}_{\rm tst}/{\rm RMSD}_{\rm trn}<1.05$.
The score
\begin{align}
 \mathcal R_{\rm fit}=\left[\frac1{N_{\rm obs}}\sum_n(z_n^{\rm th}-z_n^{\rm exp})^2\right]^{1/2}
 \label{eqn:fitscore}
\end{align}
is evaluated on all training and test observables. We save up to
ten lowest-score recorded fits from each chain and sort the combined
results by this score. The search is repeated with separately
initialized chains for each number of fitted 3NF terms.
\end{enumerate}
The search produces selected fits rather than a sample from a posterior
probability distribution. For each saved fit, we recompute the lattice energies from the
fitted coefficients and repeat the reference-variant selection.
We average the first $N$ fits in the recorded score ordering,
keeping repeated combinations of 3NF terms as separate fits.
For each lattice energy, the mean and variance are
\begin{align}
 \bar E_j^{(N)}&=\frac1N\sum_{r=1}^N E_j^{(r)},\nonumber\\
 [\bar\sigma_j^{(N)}]^2&=\frac1N\sum_{r=1}^N(E_j^{(r)}-\bar E_j^{(N)})^2+
 \frac1N\sum_{r=1}^N[\sigma_j^{(r)}]^2\,.
 \label{eqn:topN}
\end{align}
Here $N$ is the number of fits actually averaged. 
The error describes the spread of the ensemble and its mean within-fit variance, and it is not the standard error of the mean. The central tables quote $N=$~5, 50, and 100 together. The quality-versus-$p$ figures additionally compare $N=1,50,100$ to show ensemble-size sensitivity. We obtain $\bar z_n$ from the averaged energies using the same
definitions as for $z_n$. For a subset $\mathcal S$ of the energy
observables, we define
\begin{align}
 {\rm RMSD}(\mathcal S)&=\left[\frac1{|\mathcal S|}\sum_{n\in\mathcal S}(\bar z_n-z_n^{\rm exp})^2\right]^{1/2},\nonumber\\
 \delta_{\rm RMSD}&=100\,\frac{{\rm RMSD}(\mathcal S_{\rm BE})}
 {|\langle E^{\rm exp}/A\rangle_{\mathcal S_{\rm BE}}|}\,.
 \label{eqn:relative-RMSD-001}
\end{align}
Here $\mathcal S_{\rm BE}$ contains the 26 nuclear ground states,
and $|\mathcal S|$ is the number of observables in $\mathcal S$. Phase shifts are obtained from the averaged finite-volume energies by L\"uscher inversion with Gaussian propagation of their uncertainties.

\subsection{Binding energies and neutron--alpha scattering}
\label{subsec:joint}
\begin{table}[htb]
\caption{Joint fits to the 26 binding energies and twelve $n$--$\alpha$ finite-volume energies using the set of 40 terms. Relative binding-energy RMSD $\delta_{\rm RMSD}$, Eq.~(\ref{eqn:relative-RMSD-001}), and RMSDs of the $^2P_{3/2}$ and $^2P_{1/2}$ relative energies ($L=7$--$12$) are evaluated on the top-$N$ averages, Eq.~(\ref{eqn:topN}). Without fitted short-range 3NFs these values are $21.6\%$, $1.54$ and $0.98$~MeV. The spatially extended Cartesian configurations are defined in App.~\ref{app:cartesian}.}\label{tab:central-A}
\centering\footnotesize\renewcommand{\arraystretch}{1.05}
\begin{tabular}{|c||c|c|c||c|c|c||c|c|c|}\hline\hline
$p$ & \multicolumn{3}{c||}{$N=5$} & \multicolumn{3}{c||}{$N=50$} & \multicolumn{3}{c|}{$N=100$} \\
 & $\delta_{\rm RMSD}$ & $^2P_{3/2}$ & $^2P_{1/2}$ & $\delta_{\rm RMSD}$ & $^2P_{3/2}$ & $^2P_{1/2}$ & $\delta_{\rm RMSD}$ & $^2P_{3/2}$ & $^2P_{1/2}$ \\
 & (\%) & (MeV) & (MeV) & (\%) & (MeV) & (MeV) & (\%) & (MeV) & (MeV) \\\hline\hline
3 & 2.69 & 0.327 & 0.465 & 2.68 & 0.327 & 0.469 & 2.68 & 0.324 & 0.472 \\\hline
4 & 2.41 & 0.170 & 0.235 & 2.43 & 0.167 & 0.235 & 2.45 & 0.163 & 0.238 \\\hline
5 & 2.21 & 0.182 & 0.231 & 2.22 & 0.174 & 0.234 & 2.27 & 0.173 & 0.233 \\\hline
6 & 2.10 & 0.166 & 0.241 & 2.15 & 0.160 & 0.245 & 2.17 & 0.162 & 0.244 \\\hline
7 & 1.80 & 0.156 & 0.251 & 1.95 & 0.164 & 0.251 & 1.93 & 0.171 & 0.259 \\\hline
8 & 1.83 & 0.161 & 0.243 & 1.84 & 0.165 & 0.267 & 1.92 & 0.165 & 0.279 \\\hline
\hline\end{tabular}\end{table}

\begin{figure}[htb]
\centering\includegraphics[width=0.60\textwidth]{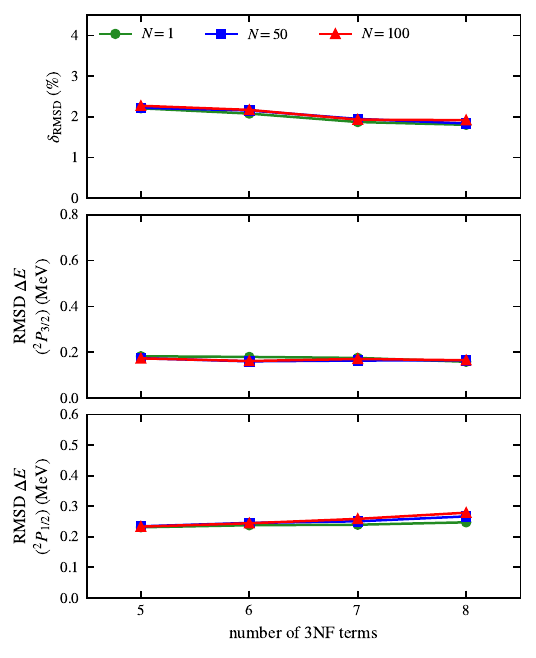}
\caption{Fit quality as a function of the number $p$ of fitted three-nucleon
terms in the main analysis, allowing the spatially extended
contact configurations. The panels show the relative binding-energy RMSD of the 26 nuclei and the RMSDs of the six finite-volume energies in each neutron--alpha channel. Curves use the first $N=$~1,~50,~100 ranked fits, where $N=1$ is the first-ranked fit.}
\label{fig:quality-A}
\end{figure}

\begin{table}[htb]
\caption{Lattice LECs for the displayed ensembles of Tab.~\ref{tab:central-A}. Listed terms occur in a first-ranked fit or in at least 40\% of one displayed ensemble. $c_{\rm first}$ is the first-ranked coefficient, $\langle c\rangle\pm\sigma_c$ is the mean and population standard deviation over top-50 fits containing that term, with its fraction in parentheses. Coefficients multiply unit-coupling lattice matrix elements of Sec.~\ref{sec:3N} and are not converted to conventional dimensionless LECs.}\label{tab:LECs-A}
\centering\footnotesize\renewcommand{\arraystretch}{1.05}
\begin{tabular}{|l||r|c||r|c||r|c|}\hline\hline
3NF term & \multicolumn{2}{c||}{$p=5$} & \multicolumn{2}{c||}{$p=6$} & \multicolumn{2}{c|}{$p=8$} \\
 & $c_{\rm first}$ & $\langle c\rangle\pm\sigma_c$ (fraction) & $c_{\rm first}$ & $\langle c\rangle\pm\sigma_c$ (fraction) & $c_{\rm first}$ & $\langle c\rangle\pm\sigma_c$ (fraction) \\\hline\hline
\csuse{C_cD_sL0} & -- & $+0.624\pm0.034$ (48\%) & $+2.3718$ & $+0.864\pm1.053$ (20\%) & -- & $+2.143\pm1.350$ (26\%) \\\hline
\csuse{C_cD_sL2} & $-0.5913$ & $-0.588\pm0.110$ (18\%) & -- & $-0.440\pm0.346$ (14\%) & -- & $-0.828\pm0.568$ (18\%) \\\hline
\csuse{C_cD_sL3} & -- & $-0.520\pm0.023$ (10\%) & $-1.1983$ & $-0.633\pm0.470$ (20\%) & $+0.4960$ & $-0.720\pm0.687$ (20\%) \\\hline
\csuse{C_cD_sNL03} & -- & $-0.006\pm0.000$ (2\%) & -- & $-0.032\pm0.037$ (16\%) & $-0.1257$ & $-0.067\pm0.051$ (18\%) \\\hline
\csuse{C_cE_sNL01} & -- & -- & -- & $+0.540\pm0.011$ (6\%) & $+0.5220$ & $+0.456\pm0.148$ (12\%) \\\hline
\csuse{C_cE_044} & $-0.5910$ & $-0.548\pm0.036$ (100\%) & $-0.5247$ & $-0.520\pm0.035$ (74\%) & $-0.4948$ & $-0.515\pm0.079$ (46\%) \\\hline
\csuse{C_cE_112} & -- & -- & $+0.0543$ & $+0.044\pm0.011$ (4\%) & -- & $+0.037\pm0.029$ (24\%) \\\hline
\csuse{C_cE_125} & -- & -- & -- & $-0.036\pm0.051$ (10\%) & $-0.0862$ & $-0.084\pm0.010$ (12\%) \\\hline
\csuse{C_cE_222} & $+0.0053$ & $+0.006\pm0.001$ (100\%) & $+0.0046$ & $+0.006\pm0.001$ (92\%) & $+0.0044$ & $+0.004\pm0.001$ (82\%) \\\hline
\csuse{C_cE_224} & -- & $+0.006\pm0.001$ (8\%) & -- & $+0.040\pm0.046$ (26\%) & $+0.0750$ & $+0.067\pm0.043$ (44\%) \\\hline
\csuse{C_cE_sL1_044} & -- & -- & -- & $-0.314\pm0.069$ (28\%) & -- & $-0.328\pm0.060$ (52\%) \\\hline
\csuse{D_3S1_01} & $-0.5123$ & $-0.469\pm0.095$ (20\%) & -- & $-0.655\pm0.000$ (2\%) & -- & $-0.489\pm0.135$ (12\%) \\\hline
\csuse{D_3P0} & -- & $-1.312\pm0.000$ (2\%) & $+2.1567$ & $+1.640\pm2.245$ (48\%) & -- & $+2.274\pm2.002$ (50\%) \\\hline
\csuse{D_3P1} & $+1.5814$ & $+1.488\pm0.155$ (100\%) & -- & $+1.716\pm1.070$ (54\%) & $+1.4293$ & $+1.896\pm1.142$ (44\%) \\\hline
\hline\end{tabular}\end{table}
Table~\ref{tab:central-A} and Fig.~\ref{fig:quality-A} show the accuracy
of the binding energies and neutron--alpha finite-volume energies in
the main fits. For the top-50 ensembles, increasing the number of
fitted 3NF terms from four to five reduces the relative binding-energy
RMSD from $\Res{A}{4}{rel}\%$ to $\Res{A}{5}{rel}\%$.
The $^2P_{3/2}$ energy RMSD increases slightly from
$\Res{A}{4}{pthree}$ to $\Res{A}{5}{pthree}$~MeV, while the
$^2P_{1/2}$ RMSD remains nearly unchanged, from
$\Res{A}{4}{pone}$ to $\Res{A}{5}{pone}$~MeV.
Improved binding energies therefore do not necessarily imply improved
neutron--alpha scattering energies. Without the fitted short-range
3NF terms, the relative binding-energy RMSD is
$\csuse{ResultAbaseline}\%$.

The interaction with five fitted 3NF terms provides a compact reference for the subsequent comparisons. With six and eight fitted terms, the binding-energy RMSDs are $\Res{A}{6}{rel}\%$ and $\Res{A}{8}{rel}\%$, respectively. The corresponding channel RMSDs and their dependence on $N=$~5, 50, 100 are given together in Tab.~\ref{tab:central-A}. Fig.~\ref{fig:quality-A} also shows the sensitivity to considering one, 50, or 100 ranked fits. For five fitted terms, changing the table ensemble size gives a binding-energy RMSD range of $\Res{A}{5}{relrange}\%$ and a $^2P_{3/2}$ energy RMSD range of $\Res{A}{5}{pthreerange}$~MeV.

\begin{figure}[htb]
\centering\includegraphics[width=0.95\textwidth]{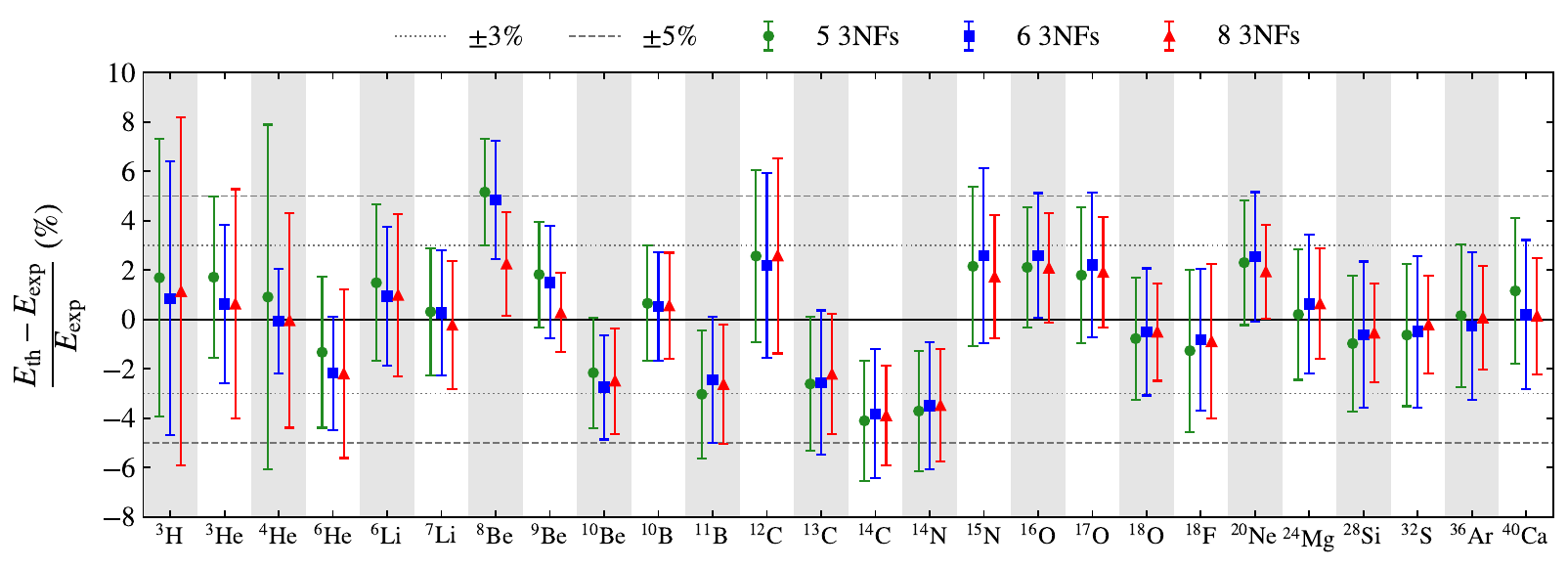}
\caption{Relative ground-state-energy deviations,
$(E_{\rm th}-E_{\rm exp})/E_{\rm exp}$, for five, six, and eight
fitted 3NF terms, using the top-50 ensembles. Error bars are obtained
from Eq.~(\ref{eqn:topN}) and divided by $|E_{\rm exp}|$, and values
are displayed in percent. The dotted and dashed horizontal lines
indicate $\pm3\%$ and $\pm5\%$, respectively. Alternating grey
backgrounds separate the nuclei.}
\label{fig:BE-reldev-A}
\end{figure}
\begin{figure}[htb!]
    \centering
    \begin{overpic}[width=1.0\textwidth]{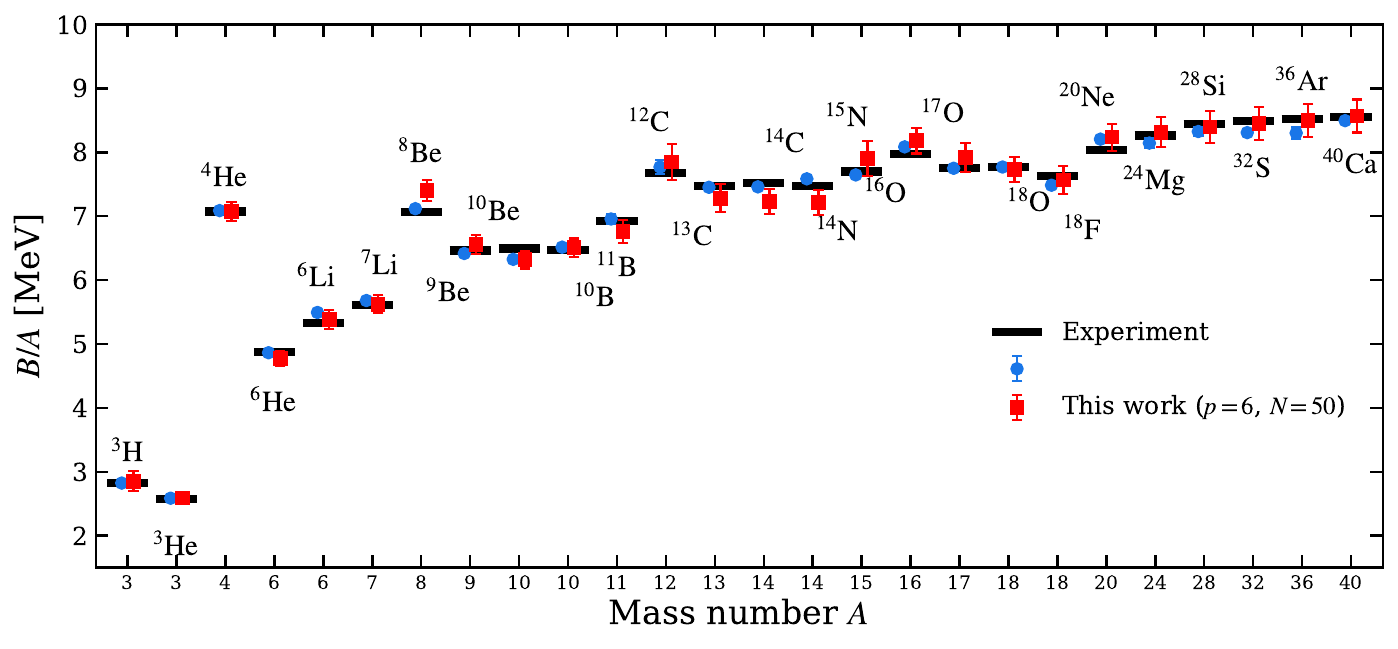}
        \put(76.0,19.4){\small Ref.~\cite{Elhatisari:2022zrb}}
    \end{overpic}
\caption{Binding energies per nucleon of the 26 nuclei. Black
horizontal bars show experiment, red squares show the top-50 ensemble
with six fitted 3NF terms from this work, and blue circles show the Wave 4
NLEFT results of Ref.~\cite{Elhatisari:2022zrb}. Error bars show one
standard deviation. $A$ gives the mass number.
}
\label{fig:BA-A}
\end{figure}

Figures~\ref{fig:BE-reldev-A} and \ref{fig:BA-A} show the binding energies nucleus by nucleus. For the ensemble with six fitted terms the largest absolute relative deviation is $\Res{A}{6}{maxdev}\%$, and the mean relative ensemble uncertainty is $\Res{A}{6}{meanerr}\%$. The global RMSD and the uncertainty bars measure different quantities. The former describes agreement of the central values with experiment, while the latter includes both reference-variant and fit-to-fit spreads.
\begin{figure}[htb!]
    \centering
    \begin{overpic}[width=1.0\textwidth]{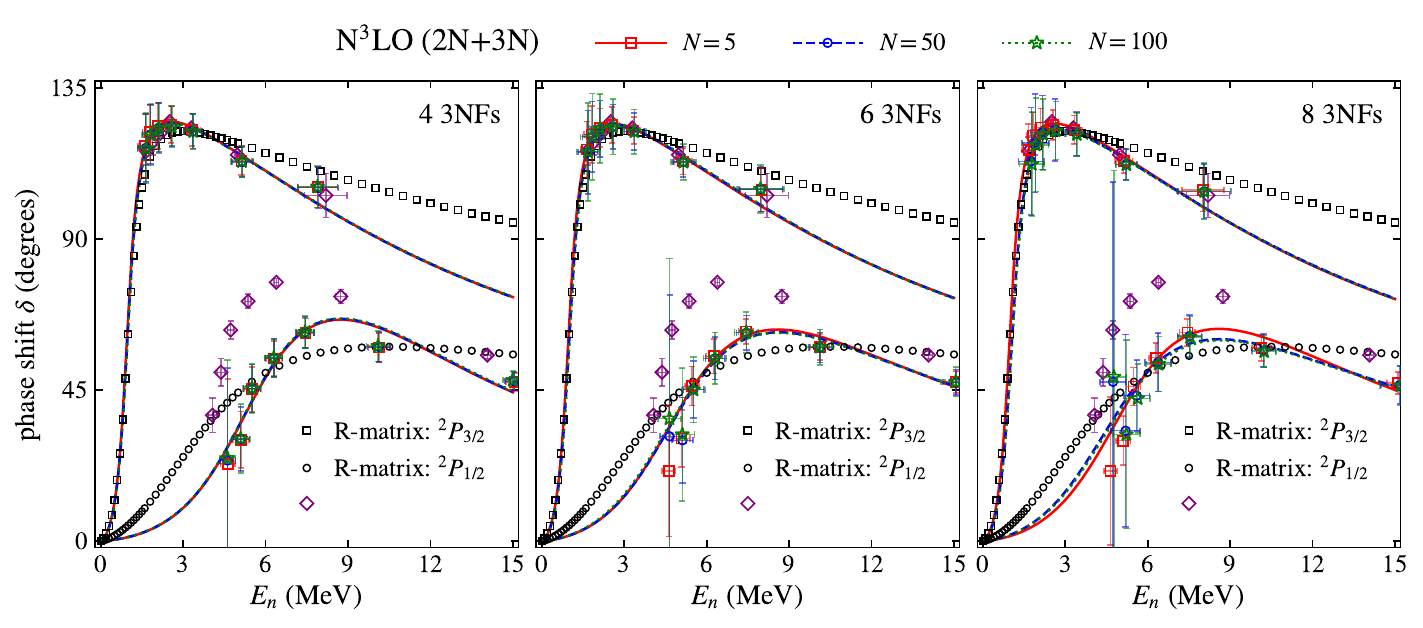}
        \put(23.5,8.3){\scriptsize Ref.~\cite{Elhatisari:2025fyu}}
        \put(54.5,8.3){\scriptsize Ref.~\cite{Elhatisari:2025fyu}}
        \put(85.7,8.3){\scriptsize Ref.~\cite{Elhatisari:2025fyu}}
    \end{overpic}
\caption{Neutron--alpha $^2P_{3/2}$ and $^2P_{1/2}$ phase shifts
for interactions with four, six, and eight fitted 3NF terms
(left to right). The red, blue, and green results use the first
$N=5,50,100$ ranked fits, respectively. Black symbols show the
empirical $R$-matrix analysis~\cite{hale2008private}, and purple
diamonds show the results of Ref.~\cite{Elhatisari:2025fyu}.
The N${}^3$LO label refers to the two-nucleon interaction supplemented
by the three-nucleon terms studied here. Lattice error bars are
propagated from the averaged finite-volume energies. Effective-range
curves are interpolation guides, which means that they do not supply an additional
uncertainty estimate.}
\label{fig:nalpha-A}
\end{figure}

Figure~\ref{fig:nalpha-A} compares the phase shifts for interactions
with four, six, and eight fitted 3NF terms. A discrepancy in the
$^2P_{1/2}$ channel remains as more terms are fitted. For the top-50
ensemble with six terms, the $^2P_{1/2}$ finite-volume energy
residuals range from $\Sci{A6ResidualMin}$ to
$\Sci{A6ResidualMax}$~MeV, changing sign between smaller and
larger boxes.

\subsection{Three-nucleon terms and their contributions}
\label{subsec:recommended}
Table~\ref{tab:LECs-A} lists the 3NF terms and fitted coefficients
for the ensembles with five, six, and eight terms. Every top-50 fit
with five terms contains the two spatially extended contact
configurations with coefficients \csuse{C_cE_044} and
\csuse{C_cE_222}, together with the partial-wave-resolved
${}^3P_1$ term \csuse{D_3P1}. The remaining two terms vary among
the fits. The common terms identify a combination favored by this
optimization, but do not isolate the contribution of any one term.
The pair partial-wave labels refer to two nucleons within the
three-nucleon interaction and should not be identified with the
neutron--alpha channels.

The tabulated coefficients use the normalization of the lattice matrix elements. Conditional coefficient averages are descriptive. In other words, observables are averages of separately evaluated fitted interactions, not predictions obtained by inserting mean coefficients into a single interaction. When more 3NF terms are fitted, several combinations give similar fit quality.

The naturalness bounds in Eq.~(\ref{eq:naturalness-bound}) constrain the optimization. Consequently, the preferred combinations of terms are conditional on this choice of bounds and on the reference interactions. Comparing different $p$ does not constitute a controlled chiral-order convergence test. The partial-wave-resolved terms provide additional spin and angular-momentum dependence, but this fit alone cannot establish their uniqueness or exclude regulator effects.

\subsection{Comparison and limitations}
\label{subsec:discussion}
The comparison using 23 terms without the spatially extended contact
configurations is given in App.~\ref{app:no-cartesian-results}.
With six fitted terms, its top-50 ensemble gives
$\delta_{\rm RMSD}=\Res{B}{6}{rel}\%$ and channel RMSDs of
$\Res{B}{6}{pthree}$ and $\Res{B}{6}{pone}$~MeV.
The main fits give smaller binding-energy and $^2P_{1/2}$ RMSDs
at larger values of $p$, while the $^2P_{3/2}$ RMSD is not
uniformly smaller. The spatially extended terms are finite-range
configurations of the regulated contact interaction.
The fixed scattering levels also differ between the two calculations,
as specified in Sec.~\ref{subsec:protocol}, and the comparison therefore
does not isolate the effect of the additional terms alone.

The reported uncertainty combines Monte Carlo errors, reference-variant spread, and variation across the selected fits. It does not include a chiral truncation estimate, lattice-spacing extrapolation, or a complete uncertainty analysis for missing three-nucleon topologies. The remaining neutron--alpha discrepancy motivates tests of additional derivative and spectator-angular-momentum structures and the omitted long-range contributions, together with independent observables such as radii. Such tests are needed before drawing conclusions about the physical origin of the residuals.

\section{Summary and outlook}
\label{sec:summary}

We have studied short-range three-nucleon forces in nuclear lattice effective field theory using an N${}^3$LO two-nucleon interaction and the fixed two-pion-exchange 3NF. The main fits combine smeared $c_D$ and $c_E$ terms, partial-wave-resolved contacts in which a nucleon pair couples to a scalar-isoscalar spectator density, and the spatially extended Cartesian contact configurations used in earlier lattice calculations. Their definitions are collected in App.~\ref{app:cartesian}, and a fit without them is retained in App.~\ref{app:no-cartesian-results}. The pair structures through $Q^2$ are considered without implying a complete interaction at their  chiral orders.

A stochastic search selects both sets of 3NF terms and fitted observables from nuclear binding energies and neutron--alpha finite-volume levels. Final observables are averages over the first 5, 50, and 100 fits in the recorded score ordering, with errors combining the between-fit variance and mean within-fit variance. For the top-50 ensemble with five fitted 3NF terms in the main analysis, the relative binding-energy RMSD is $\Res{A}{5}{rel}\%$ and the $^2P_{3/2}$ and $^2P_{1/2}$ energy RMSDs are $\Res{A}{5}{pthree}$ and $\Res{A}{5}{pone}$~MeV. The ensemble with six fitted terms gives $\Res{A}{6}{rel}\%$ for binding energies, and the corresponding result from the set of 23 terms is $\Res{B}{6}{rel}\%$. Tables~\ref{tab:central-A} and \ref{tab:central-B} quantify the dependence on the number of fitted 3NF terms and ensemble size.

For the top-50 ensembles, the binding-energy RMSD decreases as more 3NF terms are fitted,
whereas the neutron--alpha energy RMSDs do not improve monotonically.
All top-50 fits with five terms contain the spatially extended contact
configurations with coefficients \csuse{C_cE_044} and
\csuse{C_cE_222}, together with the ${}^3P_1$ pair term
\csuse{D_3P1}. Tests of additional three-nucleon contributions and
independent observables such as radii, together with truncation and
lattice-spacing studies, are needed to assess the predictive
precision of these interactions. Furthermore, extending these considerations to larger masses
might necessitate a re-analysis, as e.g. the calculation of the Sn isotopes shows~\cite{Hildenbrand:2025voq}.
One might also consider the possibility of including the nuclear matter saturation properties into the fit protocol,
as sometimes done in the continuum. In a lattice setup, such an approach was already followed in Refs.~\cite{Tong:2024jvs,Tong:2025fzv}.
Work along these lines is underway.

\acknowledgements
We are grateful for discussions with members of the Nuclear Lattice Effective Field Theory Collaboration. SE  would like to acknowledge the support provided by the Deanship of Research at King Fahd University of Petroleum \& Minerals (KFUPM) under Research Grant EC241026). We acknowledge funding in part by the European Research Council (ERC) under the European Union's Horizon 2020 research and innovation programme (EXOTIC, grant agreement No. 101018170). The work of UGM was supported in part by the CAS President's International Fellowship Initiative (PIFI) (Grant No.~2025PD0022). Computational resources provided by the Gauss Centre for Supercomputing e.V. (www.gauss-centre.eu) for computing time on the GCS Supercomputer JUWELS at J{\"u}lich Supercomputing Centre (JSC) and special GPU time allocated on JURECA-DC as well as the Oak Ridge Leadership Computing Facility through the INCITE award ``Ab-initio nuclear structure and nuclear reactions'', and partially provided by TUBITAK ULAKBIM High Performance and Grid Computing Center (TRUBA resources). Computational resources were also partly provided by the National Supercomputing Center of Korea with supercomputing resources including technical support (KSC-2021-CRE-0429, KSC-2022-CHA-0003, KSC-2023-CHA-0005).

\appendix
\section{Cartesian decomposition of the three-nucleon contact interaction}
\label{app:cartesian}

The spatially extended three-nucleon contact configurations collected in this appendix were introduced in Ref.~\cite{Elhatisari:2022zrb}, where two of them, $[V_{c_E}^{(0)}]_{114}$ and $[V_{c_E}^{(0)}]_{222}$ (denoted $V_{c_E}^{(l)}$ and $V_{c_E}^{(t)}$ there), were part of the eight-term three-nucleon interaction. The configurations considered here were already used in Ref.~\cite{Ren:2025vpe}. The fits using these configurations form the primary results in Sec.~\ref{sec:results-and-discussion}, and Appendix~\ref{app:no-cartesian-results} gives the comparison using 23 terms without them.

In Sec.~\ref{sec:3N} the locally and nonlocally smeared forms of the three-nucleon contact and OPE potentials were separated into components according to the range and strength of the smearing parameters. Here we focus on the three-nucleon contact interaction defined in Eq.~(\ref{eqn:V_cE-sL-001}) and decompose it in Cartesian coordinates into different components associated with regulator-dependent coefficients. We consider the thirteen SU(4)-symmetric configurations listed in Tab.~\ref{tab:cE-abc-config}. They fit within a cube of side length two lattice units, with pair separations no larger than $\sqrt{5}$ lattice units.
\begin{table}[ht]
	\caption{SU(4) symmetric three-nucleon contact interactions with maximum distance $\sqrt{5}$ (in lattice units).}
	\label{tab:cE-abc-config}
	\centering
	\begin{tabular}{ |c|| c | c| c | c| c | c |c | c| }
		\hline\hline				
		Coefficients  & 
		\csuse{C_cE_011} 	& 
		\csuse{C_cE_022} & 
		\csuse{C_cE_033} 	& 
		\csuse{C_cE_044} &  
		\csuse{C_cE_114} & 
		\csuse{C_cE_112} 	&
		\csuse{C_cE_123}  \\\hline
		$r_{\rm RMS}$~(fm)    & \Fixed{Radius011} 	& \Fixed{Radius022} & \Fixed{Radius033} & \Fixed{Radius044} & \Fixed{Radius033} & \Fixed{Radius022}  & \Fixed{Radius033} \\\hline
		\makecell{Shape}	&
		\makecell{\csuse{V_cE_011pic}}   &
		\makecell{\csuse{V_cE_022pic}}   &    \makecell{\csuse{V_cE_033pic}}   &
		\makecell{\csuse{V_cE_044pic}}   &   
		\makecell{\csuse{V_cE_114pic}}   &    \makecell{\csuse{V_cE_112pic}}    & \makecell{\csuse{V_cE_123pic}}\\\hline\hline
	\end{tabular}
	\begin{tabular}{ |c|| c | c| c | c| c | c |c | c| }
		
		Coefficients  	& 
		\csuse{C_cE_145} 	& 
		\csuse{C_cE_222} & 
		\csuse{C_cE_224} & 
		\csuse{C_cE_334} 	& 
		\csuse{C_cE_125}  & 
		\csuse{C_cE_235} 
		\\\hline
		$r_{\rm RMS}$~(fm)   & \Fixed{Radius145}  & \Fixed{Radius033} 	& \Fixed{Radius044} & \Fixed{Radius145} & \Fixed{Radius044} & \Fixed{Radius145} \\\hline
		\makecell{Shape}	   &    
		\makecell{\csuse{V_cE_145pic}} &
		\makecell{\csuse{V_cE_222pic}}   &    \makecell{\csuse{V_cE_224pic}}   &    \makecell{\csuse{V_cE_334pic}}   &   \makecell{\csuse{V_cE_125pic}}&
		\makecell{\csuse{V_cE_235pic}}
		\\\hline\hline
	\end{tabular}
\end{table}
The set of interactions considered here is given in Tab.~\ref{tab:cE-abc-config}. In the first four interactions, we choose the two nucleons to sit on the same lattice site and the third one is on the nearest-neighboring, the next-to-nearest-neighboring, the next-to-next-to-nearest-neighboring lattice site, and the site at a distance of two lattice units, respectively. The LECs associated with these interactions are \csuse{C_cE_011}, \csuse{C_cE_022}, \csuse{C_cE_033}, and \csuse{C_cE_044} where the superscripts indicate $d = 0$ ($R_{s_{\rm L}}=0$) and subscripts specify the squares of the absolute relative distance between the particle pairs, $|\vec{n}_{a}-\vec{n}_{b}|^2$ and $|\vec{n}_{a}-\vec{n}_{c}|^2$, $|\vec{n}_{b}-\vec{n}_{c}|^2$, respectively, with nucleon labels $a$, $b$ and $c$. In the next interaction, we choose three nucleons lying along a line with each nucleon one lattice unit apart from their neighbor(s) and the coefficient of this interaction is \csuse{C_cE_114}. For the rest of the interactions shown in Tab.~\ref{tab:cE-abc-config}, we choose the three nucleons being  on different sites and forming different triangle configurations. The interaction with the coefficient \csuse{C_cE_112} has isosceles right triangle shape, and the interactions with the coefficients \csuse{C_cE_123} and \csuse{C_cE_145} have scalene right triangle shape. In the next interaction with the coefficient \csuse{C_cE_222} the three nucleons form an equilateral triangle with side length equal to $\sqrt{2}$ lattice units, the interaction with the coefficient \csuse{C_cE_224} has isosceles right triangle configuration, and the interaction with the coefficient \csuse{C_cE_334} has a form of isosceles triangle configuration. Finally, the last two interactions with the coefficients \csuse{C_cE_125} and \csuse{C_cE_235} are forming obtuse triangle arrangements of three nucleons. The three-nucleon configurations shown in Tab.~\ref{tab:cE-abc-config}, despite the fact that nucleons are separated by a few lattice units, are still very compact. We calculate the root-mean-square radius, $r_{\rm RMS}$,
\begin{equation}
r_{\rm RMS} = a\left\{
\frac{1}{3}
\sum_{i = {a,b,c}} 
\left[
(n_{x,i} - X_{\rm CoM}^{abc})^2
+
(n_{y,i} - Y_{\rm CoM}^{abc})^2
+
(n_{z,i} - Z_{\rm CoM}^{abc})^2
\right]
\right\}^{1/2}
\end{equation}
for these configurations and give in Tab.~\ref{tab:cE-abc-config}.  Here, $X_{\rm CoM}^{abc}$, $Y_{\rm CoM}^{abc}$,
and $Z_{\rm CoM}^{abc}$ are the pertinent center-of-mass coordinates in lattice units.
We find that the $r_{\rm RMS}$ radii of these interactions are less than the Compton length of the
pion, $\lambda_\pi = 1/M_\pi \simeq \Fixed{PionLength}$~fm.
\begin{table}[ht]
	\caption{Locally smeared versions of the first four configurations of Tab.~\ref{tab:cE-abc-config}: the two nucleons on the common site are smeared up to nearest-neighbor distance.}
	\label{tab:cE-sL-abc-config}
	\centering
	\begin{tabular}{ |c|| c | c| c | c| }
		\hline\hline		
		Coefficients  &
		\csuse{C_cE_sL1_011}   & 
		\csuse{C_cE_sL1_022} 	& 
		\csuse{C_cE_sL1_033}  &  
		\csuse{C_cE_sL1_044}   \\\hline
		$R_{s_{{\rm L}}}$~(fm)  
		&   \Fixed{Spacing}   & \Fixed{Spacing} 	& \Fixed{Spacing} & \Fixed{Spacing}  \\\hline
		$r_{\rm RMS}$~(fm)  
		&   \Fixed{Radius011}   & \Fixed{Radius022} 	& \Fixed{Radius033} & \Fixed{Radius044}  \\\hline
		\makecell{Shape} &  
		\makecell{\csuse{V_cE_sL1_011pic}} &
		\makecell{\csuse{V_cE_sL1_022pic}}   & 
		  \makecell{\csuse{V_cE_sL1_033pic}}   &
		    \makecell{\csuse{V_cE_sL1_044pic}} \\\hline\hline
	\end{tabular}
\end{table}
Finally, we introduce the locally smeared form of the first four interactions given in Tab.~\ref{tab:cE-abc-config}. We consider the configurations where the two nucleons sit on the same lattice site and the third one is separated by 1-2 lattice units, and we smear the density operators of the two nucleons on the same lattice site locally only up to the nearest neighboring distance. The range of the smearing, the $r_{\rm RMS}$ radii and the illustrative figures of these interactions are given in Tab.~\ref{tab:cE-sL-abc-config}. The main motivation for introducing these interactions is to understand the influence of smearing on some of the three-nucleon configurations shown in Tab.~\ref{tab:cE-abc-config}. Explicitly, following Ref.~\cite{Ren:2025vpe}, these operators read
\begin{align}
[V_{c_E}^{(1)}]_{0bb}
={}&[c_E^{(1)}]_{0bb}\sum_{\vec n,\vec n',\vec n''}
:\rho^{(1)}(\vec n)\rho^{(1)}(\vec n')\rho^{(0)}(\vec n''):\nonumber\\
&\times\delta_{|\vec n-\vec n'|^2,0}
\delta_{|\vec n-\vec n''|^2,b}
\delta_{|\vec n'-\vec n''|^2,b}\,,\qquad b=1,\ldots,4\,.
\label{eqn:V_cE-sL-0bb}
\end{align}

\begin{table}[htb!]
\caption{Joint fits to the 26 binding energies and twelve $n$--$\alpha$ finite-volume energies using the set of 23 terms. Relative binding-energy RMSD $\delta_{\rm RMSD}$, Eq.~(\ref{eqn:relative-RMSD-001}), and RMSDs of the $^2P_{3/2}$ and $^2P_{1/2}$ relative energies ($L=7$--$12$) are evaluated on the top-$N$ averages, Eq.~(\ref{eqn:topN}). Without fitted short-range 3NFs these values are $21.6\%$, $1.54$ and $0.98$~MeV.}\label{tab:central-B}
\centering\footnotesize\renewcommand{\arraystretch}{1.05}
\begin{tabular}{|c||c|c|c||c|c|c||c|c|c|}\hline\hline
$p$ & \multicolumn{3}{c||}{$N=5$} & \multicolumn{3}{c||}{$N=50$} & \multicolumn{3}{c|}{$N=100$} \\
 & $\delta_{\rm RMSD}$ & $^2P_{3/2}$ & $^2P_{1/2}$ & $\delta_{\rm RMSD}$ & $^2P_{3/2}$ & $^2P_{1/2}$ & $\delta_{\rm RMSD}$ & $^2P_{3/2}$ & $^2P_{1/2}$ \\
 & (\%) & (MeV) & (MeV) & (\%) & (MeV) & (MeV) & (\%) & (MeV) & (MeV) \\\hline\hline
2 & 4.50 & 0.500 & 0.451 & 3.83 & 0.683 & 0.445 & 3.62 & 0.749 & 0.461 \\\hline
3 & 4.33 & 0.282 & 0.491 & 3.78 & 0.526 & 0.442 & 3.64 & 0.582 & 0.445 \\\hline
4 & 3.40 & 0.322 & 0.516 & 3.08 & 0.383 & 0.552 & 3.12 & 0.425 & 0.530 \\\hline
5 & 2.99 & 0.095 & 0.385 & 3.03 & 0.124 & 0.395 & 3.02 & 0.157 & 0.406 \\\hline
6 & 2.92 & 0.110 & 0.396 & 2.89 & 0.126 & 0.406 & 2.93 & 0.129 & 0.409 \\\hline
7 & 2.88 & 0.077 & 0.390 & 2.92 & 0.086 & 0.402 & 2.92 & 0.104 & 0.407 \\\hline
8 & 2.50 & 0.129 & 0.400 & 2.59 & 0.113 & 0.412 & 2.71 & 0.112 & 0.413 \\\hline
\hline\end{tabular}\end{table}
\section{Fits without the Cartesian contact configurations}
\label{app:no-cartesian-results}

\begin{figure}[htb]
\centering\includegraphics[width=0.55\textwidth]{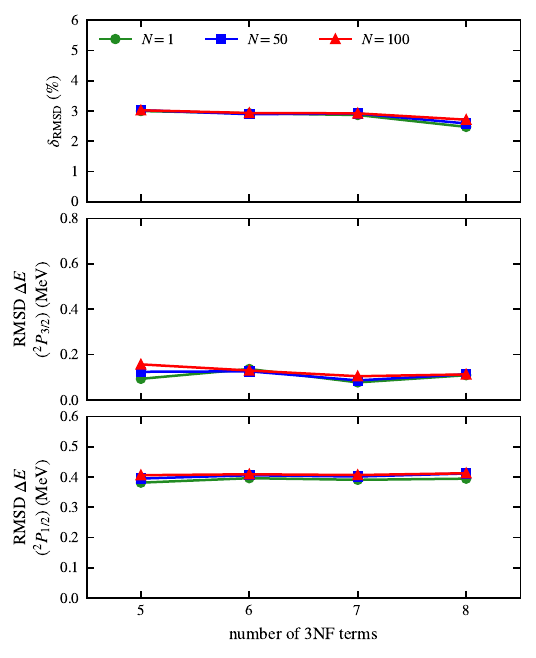}
\caption{Fit quality for the set of 23 terms without the spatially extended
contact configurations. The curves use the first $N=$~1,~50,~100 ranked fits, where $N=1$ is the first-ranked fit. The panels and metrics are defined in Fig.~\ref{fig:quality-A}.}
\label{fig:quality-B}
\end{figure}

\begin{figure}[htb]
\centering\includegraphics[width=0.9\textwidth]{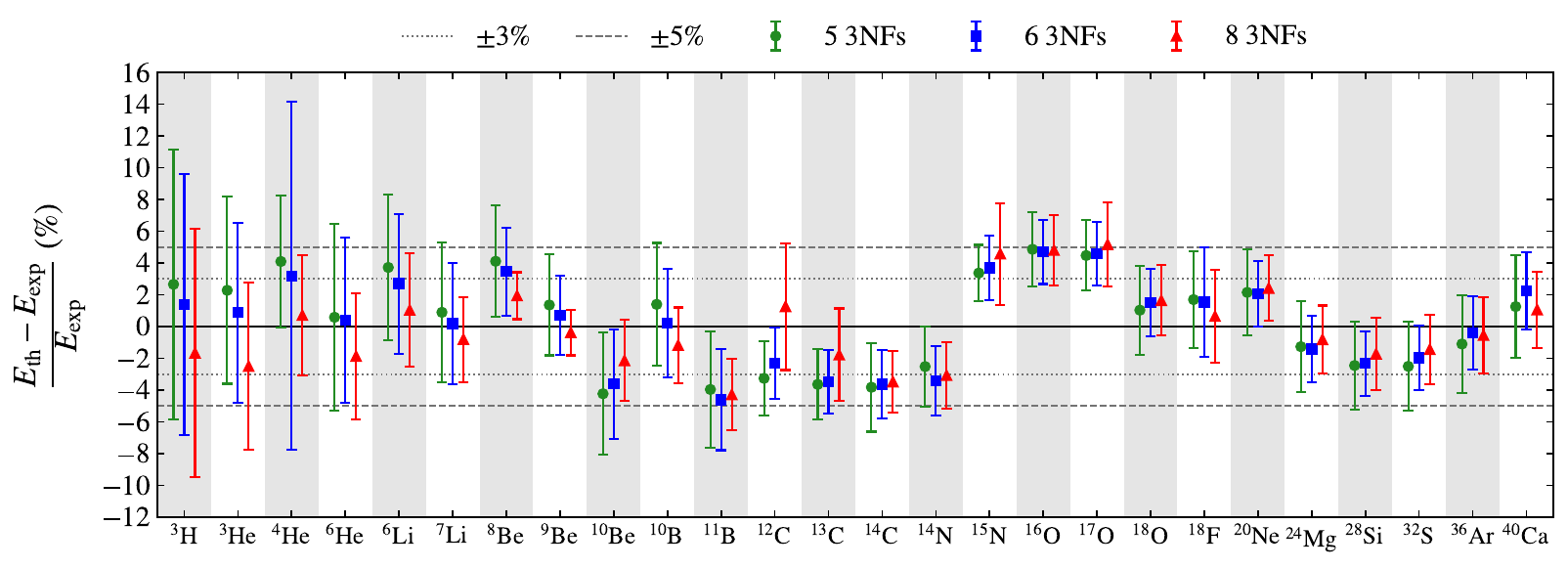}
\caption{Relative ground-state-energy deviations for five, six,
and eight fitted 3NF terms chosen from the set of 23 terms without
the spatially extended contact configurations, using top-50 ensembles.
Error bars are obtained from Eq.~(\ref{eqn:topN}) and divided by
$|E_{\rm exp}|$, and values are displayed in percent. The dotted and
dashed horizontal lines indicate $\pm3\%$ and $\pm5\%$,
respectively. Alternating grey backgrounds separate the nuclei.}
\label{fig:BE-reldev-B}
\end{figure}

\begin{figure}[htb!]
    \centering
    \begin{overpic}[width=1.0\textwidth]{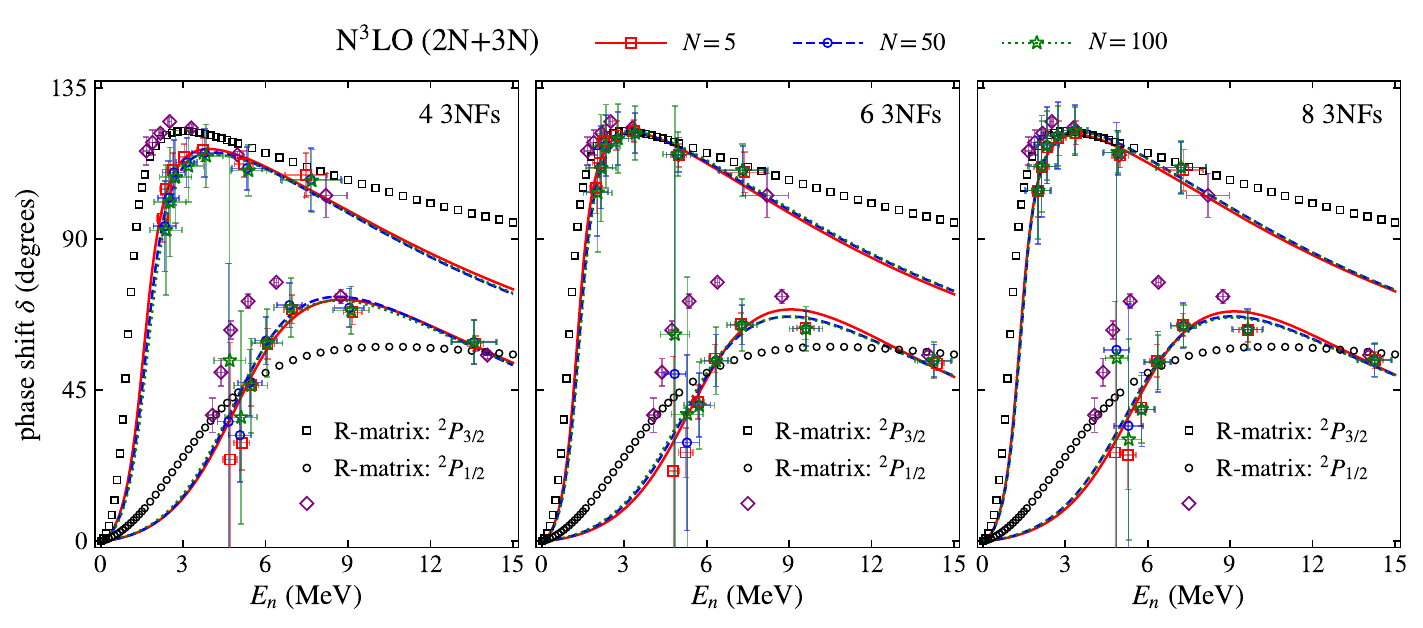}
        \put(23.5,8.3){\scriptsize Ref.~\cite{Elhatisari:2025fyu}}
        \put(54.5,8.3){\scriptsize Ref.~\cite{Elhatisari:2025fyu}}
        \put(85.7,8.3){\scriptsize Ref.~\cite{Elhatisari:2025fyu}}
    \end{overpic}
\caption{Neutron--alpha $^2P_{3/2}$ and $^2P_{1/2}$ phase shifts
for interactions with four, six, and eight fitted 3NF terms
(left to right), chosen from the set of 23 terms without the
spatially extended contact configurations. The red, blue, and green
results use the first $N=5,50,100$ ranked fits, respectively.
Black symbols show the empirical $R$-matrix analysis~\cite{hale2008private},
and purple diamonds show the results of Ref.~\cite{Elhatisari:2025fyu}.
The error bars and curves have the same meaning as in
Fig.~\ref{fig:nalpha-A}.}
\label{fig:nalpha-B}
\end{figure}

This appendix presents fits using the 23 terms formed from the smeared
$c_D$ and $c_E$ interactions and the partial-wave-resolved contacts
defined in Sec.~\ref{sec:3N}. The spatially extended contact
configurations are omitted. We use the same observable weights and
reference-variant prescription as in the main analysis, and the fixed
training levels are specified in Sec.~\ref{subsec:protocol}.
Table~\ref{tab:central-B} gives the top-5, top-50, and top-100 results
for the available numbers of fitted terms.
Figures~\ref{fig:quality-B}--\ref{fig:nalpha-B} show the same
observables as in the main text.

\begin{table}[htb]
\caption{Lattice LECs for the displayed ensembles of Tab.~\ref{tab:central-B}. Listed terms occur in a first-ranked fit or in at least 40\% of one displayed ensemble. $c_{\rm first}$ is the first-ranked coefficient, and $\langle c\rangle\pm\sigma_c$ is the mean and population standard deviation over top-50 fits containing that term, with its fraction in parentheses. Coefficients multiply unit-coupling lattice matrix elements of Sec.~\ref{sec:3N} and are not converted to conventional dimensionless LECs.}\label{tab:LECs-B}
\centering\footnotesize\renewcommand{\arraystretch}{1.05}
\begin{tabular}{|l||r|c||r|c||r|c|}\hline\hline
3NF term & \multicolumn{2}{c||}{$p=5$} & \multicolumn{2}{c||}{$p=6$} & \multicolumn{2}{c|}{$p=8$} \\
 & $c_{\rm first}$ & $\langle c\rangle\pm\sigma_c$ (fraction) & $c_{\rm first}$ & $\langle c\rangle\pm\sigma_c$ (fraction) & $c_{\rm first}$ & $\langle c\rangle\pm\sigma_c$ (fraction) \\\hline\hline
\csuse{C_cD_sL2} & -- & -- & -- & $-1.178\pm0.607$ (36\%) & -- & $-1.213\pm0.244$ (82\%) \\\hline
\csuse{C_cD_sL3} & $-1.4201$ & $-1.419\pm0.002$ (100\%) & $-1.4201$ & $-1.417\pm0.014$ (100\%) & $-1.3921$ & $-1.362\pm0.053$ (100\%) \\\hline
\csuse{C_cE_sL2} & -- & -- & -- & $-0.307\pm0.124$ (28\%) & $-0.9571$ & $-0.867\pm0.155$ (56\%) \\\hline
\csuse{C_cE_sL3} & -- & -- & $-0.1153$ & $-0.158\pm0.048$ (22\%) & -- & $-0.225\pm0.011$ (4\%) \\\hline
\csuse{C_cE_sNL01} & $+0.6845$ & $+0.684\pm0.003$ (50\%) & -- & $+0.622\pm0.064$ (8\%) & $+0.3967$ & $+0.550\pm0.110$ (18\%) \\\hline
\csuse{C_cE_sNL02} & -- & $+0.144\pm0.002$ (50\%) & $+0.1469$ & $+0.144\pm0.004$ (54\%) & $+0.1470$ & $+0.123\pm0.027$ (60\%) \\\hline
\csuse{C_cE_sNL03} & -- & -- & -- & $+0.038\pm0.010$ (28\%) & -- & $+0.026\pm0.008$ (60\%) \\\hline
\csuse{D_1S0} & -- & -- & -- & $+0.126\pm0.002$ (4\%) & -- & $+0.217\pm0.051$ (42\%) \\\hline
\csuse{D_3S1} & $-3.5804$ & $-3.576\pm0.007$ (100\%) & $-3.5759$ & $-3.483\pm0.163$ (72\%) & $-3.5133$ & $-3.486\pm0.082$ (56\%) \\\hline
\csuse{D_1S0_01} & -- & $-0.855\pm0.009$ (44\%) & -- & $-0.688\pm0.371$ (40\%) & -- & $-0.825\pm0.129$ (12\%) \\\hline
\csuse{D_3S1_01} & $-0.8497$ & $-0.870\pm0.013$ (56\%) & $-0.9403$ & $-0.905\pm0.070$ (66\%) & $-0.7671$ & $-0.779\pm0.161$ (48\%) \\\hline
\csuse{D_3P0} & $+1.9910$ & $+1.807\pm0.123$ (100\%) & -- & $+1.951\pm0.254$ (56\%) & $+6.0160$ & $+2.207\pm6.811$ (42\%) \\\hline
\csuse{D_3P1} & -- & -- & $+1.2187$ & $+1.209\pm0.372$ (46\%) & -- & $+2.846\pm1.989$ (70\%) \\\hline
\csuse{D_3P2} & -- & -- & -- & -- & $+1.6123$ & $+1.610\pm0.555$ (46\%) \\\hline
\hline\end{tabular}\end{table}


\FloatBarrier
\renewcommand{\bibfont}{\small\linespread{0.98}\selectfont}
\bibliography{References}
\bibliographystyle{apsrev}

\end{document}